%% file: main.tex
\documentclass[lettersize,journal]{IEEEtran}
\IEEEoverridecommandlockouts
\usepackage{cite}
\usepackage{amsmath,amssymb,amsfonts}
\usepackage{algorithmic}
\usepackage{graphicx}
\usepackage{textcomp}
\usepackage{xcolor}
\usepackage[inline]{enumitem}
\usepackage{subcaption}
\usepackage{easyReview}
\usepackage{array}
\usepackage{float}
\usepackage{multirow}
\usepackage{comment}
\usepackage{booktabs}
\usepackage{mathtools}

\usepackage[hidelinks]{hyperref}

\newif\ifbluehighlighttwo
\bluehighlighttwofalse  
\newcommand{\bluewraptwo}[1]{\ifbluehighlighttwo{\color{blue}#1}\else#1\fi}

\newif\ifbluehighlight
\bluehighlightfalse  
\newcommand{\bluewrap}[1]{\ifbluehighlight{\color{blue}#1}\else#1\fi}

\usepackage{tikz}
\usetikzlibrary{automata,arrows,positioning,calc }

\usepackage{orcidlink}

\newtheorem{theorem}{Theorem}
\newtheorem{lemma}{Lemma}
\newtheorem{definition}{Definition}
\newtheorem{proposition}{Proposition}
\newtheorem{remark}{Remark}
\DeclarePairedDelimiter{\norm}{\lVert}{\rVert}

\def\BibTeX{{\rm B\kern-.05em{\sc i\kern-.025em b}\kern-.08em
    T\kern-.1667em\lower.7ex\hbox{E}\kern-.125emX}}

\author{IEEE Publication Technology,~\IEEEmembership{Staff,~IEEE,}
\thanks{This paper was produced by the IEEE Publication Technology Group. They are in Piscataway, NJ.}
\thanks{Manuscript received April 19, 2021; revised August 16, 2021.}}

\usepackage[acronym,nopostdot,  style=super, nonumberlist, toc]{glossaries}
\loadglsentries{acronyms}

\usepackage{amsmath,amsfonts} 
\usepackage{graphicx}

\begin{document}

\title{Predictability of Performance in Communication Networks Under Markovian Dynamics}

\author{Samie Mostafavi \orcidlink{0000-0001-9316-0414}
,~\IEEEmembership{Member,~IEEE,}
Simon Egger \orcidlink{0009-0007-9583-6044}
,~\IEEEmembership{Member,~IEEE,}
Gy\"{o}rgy D\'{a}n \orcidlink{0000-0002-4876-0223}
,~\IEEEmembership{Senior Member,~IEEE,}
         James Gross \orcidlink{0000-0001-6682-6559}
 ,~\IEEEmembership{Senior Member,~IEEE}
\thanks{This work was supported by the European Commission through the H2020 project DETERMINISTIC6G (Grant Agreement no. 101096504).}
\thanks{ Samie Mostafavi, Gy\"{o}rgy D\'{a}n, and James Gross are with the KTH Royal Institute of Technology, Stockholm, Sweden (e-mails: ssmos@kth.se; gyuri@kth.se; jamesgr@kth.se), Simon Egger is with University of Stuttgart, Germany (e-mail: simon.egger@ipvs.uni-stuttgart.de)}
}

\maketitle

\thispagestyle{plain}
\pagestyle{plain}

\begin{abstract}

With the emergence of time-critical applications in modern communication networks \bluewrap{such as \gls{V2X} systems,} there is a growing demand for proactive network adaptation and \gls{QoS} prediction. 
However, a fundamental question remains largely unexplored: How can we quantify and achieve more predictable communication systems in terms of performance?
To address this gap, this paper introduces a theoretical framework for defining and analyzing predictability in communication systems, with a focus on the impact of observations for performance forecasting.
We establish a mathematical definition of predictability based on the total variation distance between the forecast and marginal performance distributions.
A system is deemed unpredictable when the forecast distribution, providing the most comprehensive characterization of future states using all accessible information, is indistinguishable from the marginal distribution, which depicts the system's behavior without any observational input.
This framework is applied to multi-hop systems under Markovian conditions, with a detailed analysis of Geo/Geo/1/K queuing models in both single-hop and multi-hop scenarios.
\bluewrap{Additionally, we apply the framework to a random-walk-based model of \gls{QoS} for connected vehicles experiencing changing channel conditions.}
We derive exact and approximate expressions for predictability in these systems, as well as upper bounds based on spectral analysis of the underlying Markov chains.
Our results have implications for the design of efficient monitoring and prediction mechanisms in future communication networks aiming to provide dependable services.

\end{abstract}


\input{1_introduction}
\input{2_related_works}

\input{3_system_model}
\input{4abc_predictability}

\input{4d_predictability}
\input{5ab_numerics}

\input{5cd_numerics}
\input{6_7_conclusions}
\input{8_appendix}

\bibliographystyle{ieeetr}
\bibliography{refs.bib}

\begin{IEEEbiography}[{\includegraphics[width=1in,height=1.25in,clip]{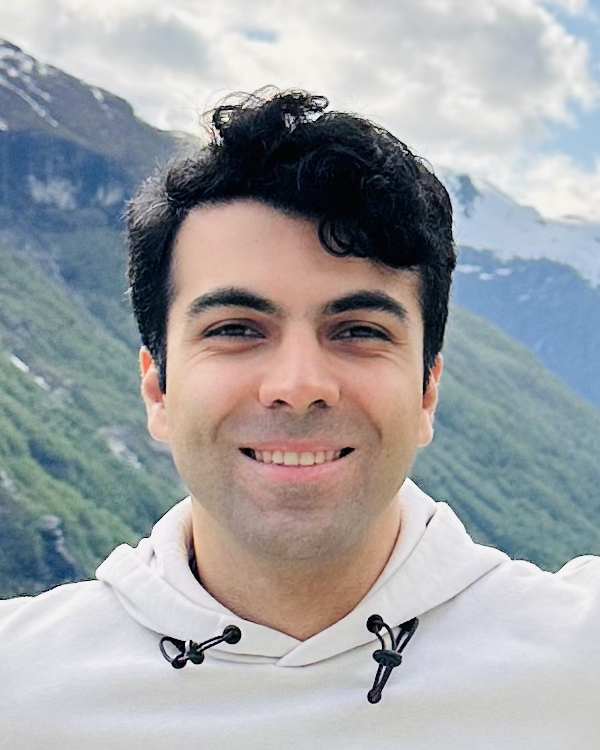}}]
{Samie Mostafavi} (Member, IEEE) received the bachelor’s degree in electrical engineering from the University of Tehran, in 2015, and the master’s degree in communication systems from KTH Royal Institute of Techonology in 2019.
He is currently pursuing the Ph.D. degree with the Department of Intelligent Systems, School of Electrical Engineering and Computer Science (EECS), KTH Royal Institute of Technology. 
He was with Ericsson AB Radio Research for a year before his doctoral studies which led to his M.Sc. thesis about vehicular positioning using 5G.
His current research is in the area of mobile edge computing and performance characterization, with a specific focus on data-driven approaches to predict delay in wireless communication systems.
\end{IEEEbiography}

\begin{IEEEbiography}[{\includegraphics[width=1in,height=1.25in,clip]{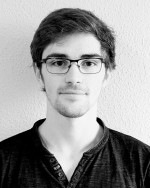}}]
{Simon Egger} (Member, IEEE) studied computer science at the University of Stuttgart and received the bachelor's degree in 2021 and the master's degree in 2024.
He is currently pursuing the Ph.D. degree at the Institute of Parallel and Distributed Systems (IPVS), University of Stuttgart.
His current research interests include adaptive and robust scheduling with formal reliability guarantees in wireless Time-Sensitive Networks.
\end{IEEEbiography}

\begin{IEEEbiography}[{\includegraphics[width=1in,height=1.25in,clip]{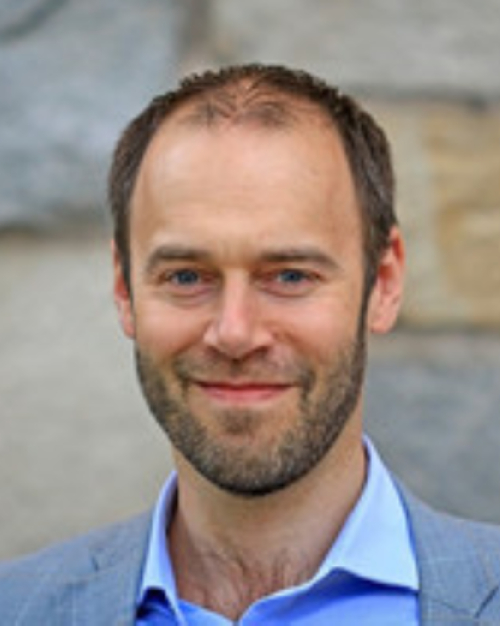}}]
{Gy\"{o}rgy D\'{a}n} (Senior Member, IEEE) received the M.Sc. degree in computer engineering from the Budapest University of Technology and
Economics, Budapest, Hungary, in 1999, the M.Sc. degree in business administration from the Corvinus University of Budapest,
Budapest, in 2003, and the Ph.D. degree in telecommunications from the KTH Royal Institute of Technology, Stockholm, Sweden,
in 2006. From 1999 to 2001, he was a consultant in the field of access networks, streaming media, and videoconferencing with BCN Ltd., Budapest. He was a visiting researcher with the Swedish Institute of Computer Science, Stockholm, in 2008, a Fulbright Research Scholar with the University of Illinois at Urbana–Champaign, Champaign, IL, USA, in 2012 and 2013, and an Invited Professor with the Swiss Federal Institute of Technology of Lausanne (EPFL), Lausanne, Switzerland, in 2014 and 2015. He is currently a Professor with the KTH Royal Institute of Technology. His current research interests include the design and analysis of content management and computing systems, game theoretical models of networked systems, and cyber-physical system security and resilience. Dr. Dan has been an Area Editor of Computer Communications 2014-2021 and the IEEE TRANSACTION ON MOBILE COMPUTING 2019-2023.
\end{IEEEbiography}

\begin{IEEEbiography}[{\includegraphics[width=1in,height=1.25in,clip]{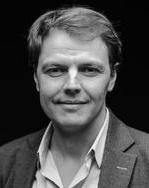}}]
{James Gross} (Senior Member, IEEE) received the Ph.D. degree from TU Berlin in 2006. 
From 2008 to 2012, he was with RWTH Aachen University, as an Assistant Professor and a Research Associate with the RWTH’s Center of Excellence on Ultra-High Speed Mobile Information and Communication (UMIC). 
Since November 2012, he has been with the Electrical Engineering and Computer Science School, KTH
Royal Institute of Technology, Stockholm, where he is a Professor of machine-to-machine communications. 
At KTH, he was the Director of the ACCESS Linnaeus Centre, from 2016 to 2019, while he is currently the Associate Director of the newly formed KTH Digital Futures Research Center, and the Co-Director of the newly formed VINNOVA Competence Center on Trustworthy Edge Computing Systems and Applications (TECoSA). 
He has authored over 150 (peer-reviewed) papers in international journals and conferences. 
His research interests are in the area of mobile systems and networks, with a focus on critical machine-to-machine communications, edge computing, resource allocation, and performance
evaluation. 
His work has been awarded multiple times, including the Best Paper Awards at ACM MSWiM 2015, IEEE WoWMoM 2009, and European Wireless 2009. 
In 2007, he was a recipient of the ITG/KuVS Dissertation Award for the Ph.D. Thesis.
\end{IEEEbiography}

\end{document}

%% file: acronyms.tex
\newacronym[plural=HITLs,firstplural=human-in-the-loop (HITL)]{HITL}{HITL}{human-in-the-loop}
\newacronym[plural=CPSs,firstplural=cyber-physical systems (CPS)]{CPS}{CPS}{cyber-physical system}
\newacronym[]{TSN}{TSN}{time sensitive networking}
\newacronym[plural=CPSs]{UE}{UE}{user equipment}
\newacronym{HARQ}{HARQ}{Hybrid-ARQ}
\newacronym{MCS}{MCS}{modulation coding scheme}
\newacronym{MLP}{MLP}{multi-layer perceptron}
\newacronym{SDR}{SDR}{software-defined radio}
\newacronym{OAI}{OAI}{OpenAirInterface}
\newacronym{SINR}{SINR}{signal to interference and noise ratio}
\newacronym{CQI}{CQI}{channel quality indicator}
\newacronym{QoS}{QoS}{quality of service}
\newacronym{PMF}{PMF}{probability mass function}
\newacronym{5GAA}{5GAA}{5G automotive association}
\newacronym{PE}{PE}{permutation entropy}
\newacronym{TV}{TV}{total variation}
\newacronym{OMM}{OMM}{observable Markov model}
\newacronym{MEC}{MEC}{mobile edge computing}
\newacronym{DVP}{DVP}{delay violation probability}
\newacronym{XR}{XR}{extended reality}
\newacronym{V2X}{V2X}{vehicle-to-everything}
\newacronym{CDF}{CDF}{cumulative distribution function}
\newacronym{RTT}{RTT}{round-trip time}
\newacronym{PTP}{PTP}{precision time protocol}
\newacronym{CDE}{CDE}{conditional density estimation}
\newacronym{GMM}{GMM}{Gaussian mixture model}
\newacronym{GMEVM}{GMEVM}{Gaussian mixture extreme value model}
\newacronym{TDD}{TDD}{time division duplex}
\newacronym[plural=PRBs,firstplural=physical resource blocks (PRB)]{PRB}{PRB}{physical resource block}
\newacronym[plural=RANs,firstplural=radio access networks (RAN)]{RAN}{RAN}{radio access network}
\newacronym{COTS}{COTS}{commercial off-the-shelf}
\newacronym[plural=MDNs,firstplural=mixture density networks (MDN)]{MDN}{MDN}{mixture density network}
\newacronym[plural=MLEs,firstplural=maximum likelihood estimation (MLE)]{MLE}{MLE}{maximum likelihood estimation}
\newacronym[plural=KL-divergences,firstplural=Kullback-Leibler divergences (KL-divergence)]{KL-divergence}{KL-divergence}{Kullback-Leibler divergence}
\newacronym[]{EVT}{EVT}{extreme value theory}
\newacronym[plural=GPDs,firstplural=generalized Pareto distributions (GPD)]{GPD}{GPD}{generalized Pareto distribution}
\newacronym[plural=PDFs,firstplural=probability density functions (PDF)]{PDF}{PDF}{probability density function}

%% file: 1_introduction.tex
\section{Introduction}

\IEEEPARstart{O}{ver} recent decades, the rapid expansion of mobile communication users and applications has intensified the demand for mission-critical services in 5G and advanced wireless networks. 
These applications—from industrial automation and remote surgery to \gls{XR}—impose stringent performance requirements, particularly regarding end-to-end latency and reliability \cite{lema2017business, elbamby2019wireless}.

To accommodate the diverse needs of these services, the community has developed tailored \gls{QoS} criteria. 
Standardized by 3GPP in 5G and continually refined for 6G, these criteria set explicit delay targets and reliability levels that limit the \gls{DVP}. 
For example, industrial control applications typically require a delay target of 10 ms and a reliability of 99.99\% \cite{5g_rel16,D11-d6g}.
\bluewrap{In \gls{V2X} scenarios—encompassing cooperative driving, collision avoidance, and autonomous navigation—networks must support latencies ranging from sub-millisecond to a few milliseconds while achieving reliability levels of 99.999\% or higher \cite{5g_v2x_sla}.

Meeting these stringent requirements is particularly challenging in networks characterized by high mobility and rapidly changing conditions, such as those encountered in \gls{V2X} communications. 
Even occasional increases in delay can lead to significant disruptions or safety hazards.
This challenge highlights both the need and the opportunity to develop accurate, proactive \gls{QoS} prediction mechanisms capable of forecasting network performance degradation, preventing accidents, and maintaining robust control loops \cite{ma_survey_2020,BARMPOUNAKIS2022109341,kousaridas_qos_2021,Tran_2023}.}

Consequently, the \gls{5GAA} introduced the concept of predictive \gls{QoS}, which refers to the mechanisms that enable mobile networks to provide notifications about predicted \gls{QoS} changes to interested consumers in advance \cite{5gaa2019making}.
The prediction of potential handovers leading to the estimated \gls{QoS} performance is the key solution for mitigating any service interruption and supporting the \gls{V2X} communication requirements.
ETSI, in the context of \gls{MEC}, has also introduced the notion of predictive \gls{QoS} support \cite{etsi_mec_002} and there is a strong focus in the development of 6G on utilizing network latency predictors, especially in the integration with \gls{TSN}, to meet the stringent requirements in scenarios such as \gls{V2X} \cite{D21-d6g,D31-d6g}.

\bluewrap{A substantial body of research has explored performance prediction for \gls{V2X} systems using machine learning techniques, examining a wide range of methods and applications \bluewraptwo{\cite{Liu2025, yazici2024, boban2022, barmpounakis2021, kim2023, torres2020}}.
In many studies, network-related metrics such as \gls{SINR}, \gls{CQI}, and network queue state served as critical inputs to throughput or delay predictive models. 
\bluewraptwo{Additionally, the notion of Age of Information (AoI), which quantifies how fresh or timely the most recently received data are, can also play a role in maintaining prediction accuracy, as outdated information may limit how accurately these models can forecast system performance~\cite{Yates2021}}.
Also, including dynamic factors like speed and location of the \gls{UE} or vehicle has proven essential for improving accuracy.
However, we contend that the confidence of performance prediction models is inherently limited by the accuracy of their inputs.
For instance, when considering dynamic factors, high \gls{UE} mobility is expected to cause rapid fluctuations in signal strength and increased variability in network delays, whereas low mobility generally results in more stable delays.
This can also be analyzed in the context of network-related metrics such as queue lengths; understanding how they evolve over time, improves our ability to assess the predictability of the performance of the network.}

Predictability in this context refers to the ability to anticipate the system's future performance based on current and historical data.
As the communication system interacts with the environment, the predictability of the system extends beyond just the immediate network parameters. 
Suppose the environmental conditions—such as interference patterns, user mobility, or traffic loads—exhibit non-random behavior and can be forecasted with reasonable accuracy. In that case, this adds a new dimension to the system's overall predictability. 
Therefore, it is essential to understand how a communication system can be considered predictable and to identify the factors that enhance or undermine this predictability.

Inspired by the above questions, in this paper, we establish a mathematical definition of predictability based on information theory concepts and a probabilistic forecasting setup.
We introduce this definition and examine its consequences within a multihop communication system model where the conditions are governed by a Markov chain.
We study both exact and approximate derivations of predictability within these abstract systems, such as a queueing system, for end-to-end delay prediction.
Queuing systems can effectively model large-scale effects in communication networks.
Our primary contributions are summarized as follows:

\begin{itemize}
    \item The paper introduces a mathematical definition of predictability based on probabilistic forecasting problem.  
    A system is considered unpredictable if the probability distribution of performance conditioned on the observations (forecast distribution) is not different from the marginal distribution of performance.
    This is done using the total variation distance between the forecast and marginal distributions to quantify the degree of predictability.

    \item We applied this framework in an abstract model with multihop systems where their performance is modulated by Markov chains and derive both exact and approximate measures of predictability.
    Our analysis explored the impact of system observability, state aggregation, and delayed observations on predictability.

    \item We utilized spectral analysis techniques to understand how the dynamics of Markovian conditions and the randomness of their transitions influence predictability.
    We derive upper bounds on predictability in Markov-modulated systems via spectral analysis and the spectral gap of the Markov chain.

    \item \bluewraptwo{ We apply our predictability derivations to a simplified queueing model, focusing on a Geo/Geo/1/K system in both single-hop and multi-hop settings. The primary performance metric is the end-to-end delay, and we also address packet loss caused by blocking. In each case, the queue length serves as the predictive feature of interest.}
    The study includes exact and approximate derivations of predictability for this queuing system.

    \item \bluewraptwo{Deriving closed-form expressions for the total variation distance between forecast and marginal distributions in Markov-modulated processes poses significant mathematical challenges. In this work, we introduce a spectral-based upper bound on predictability that exploits the Markov chain’s mixing behavior, and we further develop an approximation method for the Geo/Geo/1/K queue. These techniques reduce computational overhead while retaining analytical clarity, thereby facilitating tractable performance and predictability studies in a broad range of networking scenarios.}

    \item \bluewrap{To extend this analysis to vehicular networks, we apply our framework to evaluate the predictability of network throughput under a random walk model of \gls{CQI}, capturing how connected vehicles experience changing channel conditions.}

    \item Finally, by numerical evaluations, we showcased the utility of our predictability analysis in both queuing and random walk setup.
    The evaluations investigate the trade-offs between predictability and observability in single and multi-hop queueing systems.
    We demonstrated the accuracy of our proposed predictability measures and approximations by comparing them with exact calculations.

\end{itemize}

The rest of the paper is organized as follows: Section 2 presents an overview of the existing work in the literature on predictability. 
Section 3 presents the system model and predictability definitions, Section 4 presents the predictability analysis in Markov-modulated systems in general, while Section 5 contains the numerical evaluations and assessment of results.
Finally, Section 5 concludes this work.

%% file: 2_related_works.tex
\section{Related Works}

Researchers have developed various measures to assess the predictability of complex dynamic systems, all of which focus on predicting specific single or multi-dimensional critical metrics. 
These works can be categorized based on the following criteria: the comprehensiveness of the predictability definition, whether they analyze only the target metric or incorporate side information, and whether predictability is derived empirically or through a model-driven approach.

At the most basic level, there are works that define predictability in a setup where point prediction of a univariate metric is desired. 
Autocorrelation which evaluates how well a time series correlates with its own time-lagged versions, is a widely used indicator of predictability in finance \cite{lim_are_2013}. 
\Gls{PE} is a metric employed in fields like ecology and physics to gauge predictability \cite{bandt_permutation_2002, garland_model-free_2014, pennekamp_intrinsic_2019}.
\Gls{PE} measures time series complexity by quantifying the diversity and frequency of ordinal patterns formed by subsequences of the series.
A higher permutation entropy indicates more diverse motifs, leading to lower predictability. 
This indicator is mostly empirically evaluated and has been applied to various domains, including predicting infectious disease outbreaks \cite{scarpino_predictability_2019}, and ecological systems \cite{pennekamp_intrinsic_2019}.
Abeliuk et al. explore the relationship between predictability and sampling in partially observed systems.
They evaluate multiple predictability measures such as \gls{PE} and autocorrelation in outbreaks of infectious diseases, digital forum interactions, and software development projects \cite{abeliuk_predictability_2020}.

Besides autocorrelation and \gls{PE}, another common approach in defining predictability is to find the lower bound on prediction error.
In some works, the empirically determined entropy of the target metric is incorporated in Fano’s inequality to derive the bound on error.
Using this approach Li et al. investigate the limits of predictability for urban vehicular location and staying time \cite{li_limits_2014}.
More information-theoretic measures of predictability have inspired numerous studies.
For instance, the studies on the predictability of human mobility and conversation use random entropy, uncorrelated entropy, and conditional entropy to measure predictability \cite{song_2010,taro_2011}.

More advanced studies in this category incorporate side information or covariates into the predictability model. 
As a result, the selected predictability measure is typically relative entropy, conditional entropy, or mutual information between the side information and the target metric.
Bialek et al. define predictive information as the mutual information between past and future states of a time series as a criterion for predictability \cite{bialek_predictability_2001}.
Haven et al. use a Gaussian ensemble prediction to assess relative entropy for predictability analysis \cite{haven2005}. 
More recently, Li et al. apply conditional
entropy to traffic forecasting, highlighting the limitations of traffic predictability \cite{li_2022}.
Also, Fang and Lee explore the predictability in the context of regression problems in machine learning.
They analyze predictability in this context using the conditional entropy between the features and the label and introduce lower and upper bounds for it \cite{fang2024}.
Due to the incorporation of side information, these predictability analysis are considered more comprehensive in our perspective.

In atmospheric sciences literature, where the probabilistic forecasting of the weather systems is of interest, the information-theoretic definition of predictability with side information is gaining more attention \cite{delsole_predictability_2004,krishnamurthy_predictability_2019}.
In a series of papers, DelSole et al. define predictability as the difference between forecast and climatological distributions (marginal distribution), with the latter serving as the baseline for probabilistic information about the metric where no side information is incorporated.
Various information-theoretic measures have been proposed and analyzed to quantify predictability in systems with linear stochastic models and Gaussian processes.
\cite{delsole_predictability_2007}.

Among the works that are related to networking and communication systems, Ding et al. examine the predictability of radio spectrum state dynamics using Fano's inequality.
Their findings show high predictability in real-world spectrum measurements, which has significant implications for cognitive radio networks and 5G spectrum sharing \cite{ding_limits_2015}.
In another paper, Sihai et al. analyze voice traffic predictability in wireless networks using entropy, which captures temporal correlations in call detail records. Their study establishes a theoretical upper bound on the accuracy and demonstrates that using N-order Markov models \cite{sihai2020}.
Jing et al. study structural predictability in directed networks, linking it to controllability. They define predictability empirically through link prediction accuracy, using a leave-one-out ranking method. Unlike entropy-based approaches, their work offers no theoretical bounds but highlights the role of network structure in predictability \cite{jing2022}.

Despite the growing focus on quality of service prediction in communication systems, comprehensive predictability analyses are both necessary and, to the best of our knowledge, largely nonexistent.
This work addresses this gap by adopting a thorough predictability model and demonstrating its implications for communication system performance. 
Drawing inspiration from the predictability definition used in atmospheric science, which is an information-theoretic framework well-suited for probabilistic forecasting, our definition compares the conditional distribution of performance given the side information or observations with its marginal distribution.
Such definition offers great capabilities not only in applying it to various dynamical systems, but also we show how it can model imperfect observability.

\bluewrap{For the choice of predictability measure, no single measure is universally optimal; each measure provides unique insights depending on the specific system and the goals of the analysis.
Autocorrelation effectively captures linear dependencies in time-series data, while permutation entropy excels at quantifying non-linear structure in univariate signals. 
Entropy-based measures (e.g., cross entropy or entropy difference) provide a broad information-theoretic perspective—especially valuable when side information is involved—by quantifying the “surprise” or extra coding cost when using one distribution (forecast) in place of another (marginal). 
Entropy-based measures are highly sensitive to low-probability mismatches but are not metrics, as they generally lack symmetry and not satisfy the triangle inequality. 
In contrast, \gls{TV} distance is a true metric that directly measures the maximum difference between two distributions, remains bounded even if one distribution assigns negligible probability to events the other deems likely, and is subadditive.
Subadditivity proves particularly useful in “systems of systems” (e.g., multi-hop communication networks) and Markovian settings: the overall discrepancy is no greater than the sum of discrepancies in each subsystem.
Meanwhile, entropy-based measures offer complementary insights into average-case information loss, underscoring that the choice of measure should reflect whether one prioritizes worst-case system-level control (favoring \gls{TV} distance) or average-case informational properties (favoring entropy-based approaches) \cite{cover1999elements}.}

%% file: 3_system_model.tex
\section{System Model and Problem Formulation} \label{sec:system_model}

Forecasting performance in a communication network becomes more manageable and precise by breaking down the system into independent modules and modeling each individually allows for more efficient proactive adjustments, as it enables more precise identification of performance variations or declines \cite{mostafavi2024edaf}.
These components can correspond to tandem data transmission links or, within a single link, to multiple processes that sequentially handle data bits to ensure reliable and efficient communication.
Furthermore, the condition or state of each subsystem can be monitored and, according to domain knowledge, correlated with performance.
Before examining the system model and its predictability definition in detail, the following subsection provides a concise overview of the main parameters and their associated notations.

\bluewrap{ 
\subsection{Notations} \label{sec:notations}
In this subsection, we summarize the main notations and symbols used throughout the paper to facilitate clarity:
\begin{itemize}
    \item \(\Pr(\cdot)\): Probability distribution. For instance, \(\Pr(Z_n)\) is the probability distribution of the random variable \(Z_n\).
    \item \(\Pr(A \mid B)\): Conditional probability of event \(A\) given event \(B\).
    \item \(\lvert \mathcal{X} \rvert\): Cardinality of the set \(\mathcal{X}\), i.e., the number of elements in \(\mathcal{X}\).
    \item \(\|p - q\|_{\text{TV}}\): \Gls{TV} distance between distributions \(p\) and \(q\).
\end{itemize}
Moreover, Table \ref{tab:notation_table} provides brief explanations of the key symbols used throughout the paper.
} 
\begin{table}[!ht]
    \centering
    \caption{\bluewrap{Key Model Parameters and Their Meanings}}
    \bluewrap{
    \label{tab:notation_table}
    \begin{tabular}{|l|l|}
    \hline
    \textbf{Symbol} & \textbf{Meaning} \\ \hline
    \(n\)                                   & Time slot index (discrete-time) \\ \hline
    \(m\), \(M\)                            & Subsystem index and number of subsystems (hops) \\ \hline
    \(Z_n\) / \(Z_n^{(m)}\)                 & Performance metric r.v.\ at time slot \(n\) (subsystem \(m\)) \\ \hline
    \(\mathcal{Z}\)                         & Space of performance r.v.\ \(Z_n\) \\ \hline
    \(X_n\) / \(X_n^{(m)}\)                 & Condition r.v.\ at time slot \(n\) (subsystem \(m\)) \\ \hline
    \(\mathcal{X}\) / \(\mathcal{X}^{(m)}\) & Space of condition r.v. (size \(K\) or \(K^{(m)}\)) \\ \hline
    \(\mathcal{A}\) / \(\mathcal{A}^{(m)}\) & Space of aggegated conditions r.v. (size \(\bar{K}\) or \(\bar{K}^{(m)}\)) \\ \hline
    \(O_n\)         & Condition observation r.v.\ at time slot \(n\)\\ \hline
    \(O_{0:n}\)     & Collection of condition observations r.v. up to time $n$\\ \hline
    \(L\)           & Forecast lead time (number of slots ahead) \\ \hline
    \(d\)           & Observation delay (in slots) \\ \hline
    \(r_y(z)\)      & PMF of the posterior distribution \\ \hline
    \(P\)           & Markov chain's transition probability matrix \\ \hline
    \(\pi\)         & Markov chain's stationary distribution \\ \hline
    \(\alpha,\mu\) & Arrival and service probabilities \\ \hline
    \(\rho = \alpha / \mu\)  & Utilization factor of the queue \\ \hline
    \end{tabular}
    }
\end{table}

\begin{figure}
    \centering
    \begin{tikzpicture}[node distance=1cm, auto]
    \node (box1) [draw, rectangle, minimum width=1cm, minimum height=1cm] {Subsystem 1};
    \node (label1) [above of=box1, node distance=1cm] {$Z_n^{(1)}$};
    \node (label1b) [below of=box1, node distance=1cm] {$X_n^{(1)}$};
    
    \node (box2) [draw, rectangle, minimum width=1.9cm, minimum height=1cm, right of=box1, node distance=2.5cm] {$\cdots$};
    \node (label2) [above of=box2, node distance=1cm] {$\cdots$};
    \node (label2b) [below of=box2, node distance=1cm] {$\cdots$};
    
    \node (box3) [draw, rectangle, minimum width=1cm, minimum height=1cm, right of=box2, node distance=2.5cm] {Subsystem $M$};
    \node (label3) [above of=box3, node distance=1cm] {$Z_n^{(M)}$};
    \node (label3b) [below of=box3, node distance=1cm] {$X_n^{(M)}$};
    
    \draw[->] (box1) -- (box2);
    \draw[->] (box2) -- (box3);

    \node (circle) [below of=label2b, node distance=1.5cm] {$O_n$};
    
    \draw[->, dashed] (label1b) -- (circle);
    \draw[->, dashed] (label2b) -- (circle);
    \draw[->, dashed] (label3b) -- (circle);
    
    \end{tikzpicture}
    \caption{Multi-hop communication system model with observable measures being conditions.}
    \label{fig:sysmodel}
\end{figure}

\begin{figure}
    \centering
    \begin{tikzpicture}
      \tikzset{node style/.style={circle, draw, minimum size=0.8cm, outer sep=0, inner sep=0, text width=0.8cm, align=center}}
      \node[node style] (Xn1) at (-1.75, 1.5) {$X_{n-1}^{(m)}$};
      \node[node style] (Xn) at (0, 1.5) {$X_{n}^{(m)}$};
      \node[node style] (Xn1p) at (1.75, 1.5) {$X_{n+1}^{(m)}$};
      \node[node style] (Zn1) at (-1.75, 0) {$Z_{n-1}^{(m)}$};   
      \node[node style] (Zn) at (0, 0) {$Z_{n}^{(m)}$};  
      \node[node style] (Zn1p) at (1.75, 0) {$Z_{n+1}^{(m)}$};
      \node at (-3, 1.5) {$\dots$};
      \node at (3, 1.5) {$\dots$};   
      \draw[->, thick] (Xn1) -- (Xn);
      \draw[->, thick] (Xn) -- (Xn1p);
      \draw[->, thick] (Xn1) -- (Zn1);
      \draw[->, thick] (Xn) -- (Zn);
      \draw[->, thick] (Xn1p) -- (Zn1p);
    \end{tikzpicture}
    \caption{Temporal dependencies in the performance model for each subsystem $m$, underscoring the conditions' Markov chain}
    \label{fig:markovmodel}
\end{figure}

\subsection{Multi-hop Communication System Model}
We consider a communication system that consists of $M$ independent subsystems or hops as shown in Figure \ref{fig:sysmodel}.
As packets traverse this multi-hop network, in time slot $n$ (time is slotted), subsystem $m$ exhibits a performance measure denoted by $Z^{(m)}_n$ which is a random variable over the discrete space $\mathcal{Z}$, for example, it can denote the delay of subsystem $m$.
In this case, the end-to-end delay of the system (without packet drops) is obtained via $Z_n = \sum_{m=1}^{M} Z_n^{(m)}$.
\bluewrap{This multi-hop structure can, for instance, represent an autonomous vehicle scenario in which sensor data must first traverse a wireless uplink from the vehicle to a remote control server (hop 1), undergo processing to generate control commands (hop 2), and then return via a wireless downlink (hop 3) to the vehicle. Each hop introduces delay, and low end-to-end latency is crucial for guaranteeing system safety.}

Moreover, each subsystem has an internal state or condition measure, denoted by $X^{(m)}_n$ which influences its delay. 
For example, the compute subsystem may experience congestion, while a wireless link hop might be subject to unfavorable fading conditions, leading to significant packet delays.
These states lie in a discrete finite space $\mathcal{X}^{(m)}$ of size $K^{(m)}$.
Collectively, all subsystem state variables form the overall system state, $X_n = \{ X^{(1)}_{n},...,X^{(m)}_{n},...,X^{(M)}_{n} \}$  defined over a discrete space $\mathcal{X}$ with size $K$.

\bluewrap{We assume a stationary stochastic dependency between the state and the performance of each subsystem, i.e.,
\begin{equation} \label{eq:assump1}
    \forall L, \Pr(Z_{n+L} = z | X_{n+L} = x) = \Pr(Z_n = z | X_n = x).
\end{equation}
This definition of stationarity means that the conditional distribution of performance given the state does not change over time, and it is fundamental to able to learn from past data and predict future performance.
We argue that stationarity is a reasonable assumption that holds over extended periods of time. We denote the conditional distribution of the performance by $r_x(z)=\Pr(Z_n=z \mid X_n=x)$.

We consider that each subsystem's state can be observed, and observations can be transmitted to a central server.
This assumption is motivated by advancements in next-generation wireless technologies (e.g., 5G, 6G), which emphasize real-time monitoring of network conditions to enhance \gls{QoS} \cite{ma2020survey}. 
} 
We denote the observations of subsystem states up to time $n$ by \(O_{0:n}\).
Depending on the scenario, these observations may represent all, part, or an imperfect version of the true subsystem states.
\bluewrap{This assumption also recognizes that as monitoring capabilities grow, they inevitably introduce overhead into the system, requiring optimization of their resource usage. 
For example, \gls{SINR} measurements and mobility data may be collected and uploaded via the wireless uplink to a monitoring server for decision-making. 
These monitoring links themselves can involve delays and lossy compression affecting accuracy.}

We consider three different observability defects in our model: first, delayed observations are particularly crucial when a network of wireless remote nodes is involved in the packet route.
In this scenario, monitoring information would arrive with a delay to the prediction and adaptation point, i.e., $O_n = X_{n-d}$.
In the second case, we may choose to ignore certain subsystems and not observe their state, i.e., $O_n \subset X_n$ due to the overhead that the monitoring task introduces.
The third case, which also concerns overhead, mandates aggregated states.
In this scenario, we assume states of $\mathcal{X}$ are aggregated to $\mathcal{A}$, i.e., there exists a surjective map $\phi : \mathcal{X} \rightarrow \mathcal{A}$ with $| \mathcal{A} | \leq | \mathcal{X} |$ (or $\bar{K} \leq K$ as $| \mathcal{A} | = \bar{K}$).
Then, the observations are coduncted on the aggregated states $O_n = \phi(X_n)$.

\subsection{ \bluewrap{Markovian Model of System State} }

We model the evolution of subsystem conditions over time as a first-order Markov process. This assumption posits that the future state of a subsystem, denoted by $X_n$, depends only on its current state, $X_{n-1}$, and the associated transition probabilities.  Formally, this means that the probability of transitioning to a future state, $X_n = x_i$, given the entire history of states up to time $n-1$, is equivalent to the probability of transitioning to that state given only the current state:
\begin{equation}
    \Pr(X_n = x \mid X_0, X_1, \dots, X_{n-1}) = \Pr(X_n = x \mid X_{n-1}).
\end{equation}
These transition probabilities, which govern the dynamics of the system's state changes, encapsulate how likely a subsystem is to move from one condition to another in a single time step. The Markov property simplifies the analysis by abstracting the system's history; the past only influences the future indirectly, through its contribution to the present state.

\bluewrap{This Markovian model offers a tractable yet informative model for the evolution of network states in dynamic environments like vehicular and wireless networks.
While real-world network behavior is complex, the Markov property allows us to focus on the most recent influences on a subsystem's state and the typical patterns of change from that state. 
In high-mobility scenarios, the current location and surrounding environment of a \gls{UE} often play a dominant role in determining short-term network states.
Markovian models were used in works focusing on various aspects of network analysis and optimization, including vehicular network performance, channel allocation, and resource management \cite{ramezani2021, alshanyour2009, si2010}.
While acknowledging the inherent limitations of this model, particularly regarding long-term dependencies, the Markov assumption provides a valuable balance between model accuracy and computational feasibility, enabling efficient analysis and real-time adaptation in dynamic network settings.}

A direct result of the Markovian assumption and Equation \ref{eq:assump1} on the temporal dependencies in the performance model, as shown in Figure \ref{fig:markovmodel} is that 
\begin{equation}
    \Pr(Z_n \mid Z_{0:n-1}, X_{0:n}) = \Pr(Z_n \mid X_n).
\end{equation}
This system forms an \gls{OMM} where the sequence $\{ X_0, X_1, ..., X_n \}$ is a discrete-time, time-homogeneous Markov chain with finite state space $\mathcal{X}$.
The transition probability from state $x$ to state $y$ is denoted by $P(x,y) = \Pr(X_{n+1} = y \mid X_{n} = x)$ which forms the transition probability matrix $P$.
$L$ step state transition probability is derived via $P^{L}(x,y)$.
Moreover, we assume that $X$ is an irreducible and aperiodic Markov chain, therefore it has an stationary distribution $\pi$ that state probabilities converge to for large lead times.
Then, for all $x, y$ we have $P^{L}(x,y) \rightarrow \pi(y)$ as $L \rightarrow \infty$.
Also, we assume the chain is reversible, which is an assumption satisfied by all random walks on undirected graphs, as well as many other Markov chains such as queues:
\begin{equation}
    \forall (x,y)\in\mathcal{X}^2, \text{   } \pi(x) P(x,y) = \pi(y) P(y,x).
\end{equation}
Reversibility allows applying spectral analysis on the speed of convergence that we use later on.
In reversible chains, the spectral theorem guarantees that $P$ admits $K = |\mathcal{X}|$ real eigenvalues, which can be ordered as
\begin{equation}
    1 = \lambda_1 \geq \lambda_2 \geq ... \geq \lambda_K > -1,
\end{equation}
and there is an orthonormal basis of real-valued eigenfunctions $(f_j)_{j \leq K}$ corresponding to these eigenvalues with respect to the inner product $\langle\cdot,\cdot\rangle_{\pi}$ \cite{Winkler2003}.
Closed-form solutions for eigenvalues and eigenfunctions exist for numerous basic and well-organized Markov chain models, including random walks on circular structures, multi-dimensional cubes, and various graph types.

We will also examine more specialized Markov chains, particularly the Geo/Geo/1/K queue, which represents the discrete-time counterpart of the M/M/1/K queue.
In this case, the packets arrive according to a Bernoulli process and the probability of arriving during a time slot is $\alpha = 1-\bar{\alpha}$.
Service times are geometrically distributed, and the probability of completion of a service during a slot is $\mu = 1-\bar{\mu}$.
The utilization factor is denoted by $\rho = \alpha / \mu$.
The performance metric $Z_n$ will be the sojourn time of the queue, and the queue size will be the measurable state of the system $X_n$.
\bluewraptwo{As the number of states in our analysis should be finite, we consider a maximum length of $K$ for the queue which in actual implementations corresponds to the system’s buffer size. Accordingly, $K^{(m)}$ would denote the buffer size at hop $m$. Because the queue is of finite length, packets arriving when the buffer is full are dropped, making this a lossy system,  underscoring the importance of considering the buffer size when analyzing packet delay or packet loss \cite{Balakrishnan2021}.}

\bluewraptwo{
Furthermore, to explore the predictability of network performance in connected vehicles, we consider that these networks often exhibit variable channel quality (e.g., \gls{CQI}) due to fluctuations in mobility, interference, and propagation conditions. 
While exact modeling of these variations can be highly complex, a random walk process provides a tractable approximation for the sequential dependency observed in the \gls{CQI}.
The \gls{CQI} is a standard metric in cellular networks such as LTE and 5G, indicating link quality between a base station and the \gls{UE}.
\Gls{CQI} variations can be substantial or minimal, depending on factors such as mobility, and are important because the connection’s throughput depends on the \gls{CQI} \cite{Raca2018}. } 
In this model, states are represented within a finite one-dimensional space of size $K$. At each step, the state either remains unchanged with a certain probability or transitions to a neighboring state with equal likelihood (except for at the reflecting boundary). The process maintains a uniform stationary distribution, meaning that for any state $y$, the probability $\pi(y)$ is consistently $1/K$.
Each \gls{CQI} level corresponds to a particular throughput distribution, and transition probabilities between \gls{CQI} levels we estimate from real-world mobile vehicle connection quality measurements \cite{Raca2018}.
Random walk models offer a simplified yet effective way to represent the stochastic nature of vehicular movement. These models have had numerous applications in the study and analysis of vehicular mobility within connected vehicle environments \cite{bai2004survey, michelini2008, ajaz_simulation_2024}.

\subsection{\bluewrap{Predictability Model and Definitions}}

The primary objective of any performance predictor in this setting is to capture the relationship between the observations and future performance. 
The probability distribution of the performance metric, $L$ timeslots into the future, based on the observations collected up to the present moment, is the \textit{forecast distribution}:
\begin{equation}
    \Pr(Z_{n+L} \mid O_{0:n} = o_{0:n}).
    \label{eq:fc}
\end{equation}
The distribution of performance right after observations become available denoted by $\Pr(Z_n \mid O_{0:n})$ is the \textit{posterior distribution} \cite{Reich_Cotter_2015}.
The \textit{marginal distribution} of performance (or prior distribution) denoted by $\Pr(Z_n)$ is related to the posterior distribution by the chain rule as it is obtained by integrating out all observations:
\begin{equation}
    \Pr(Z_{n}) = \sum_{o_{0:n}} \Pr(Z_{n} \mid O_{0:n} = o_{0:n} ) \Pr( O_{0:n} = o_{0:n} ).
\end{equation}
The marginal distribution constitutes a baseline for the information content about performance, disregarding all state observations.
\bluewrap{This distribution can be derived by randomly sampling performance values and remains stationary because: (i) the conditional distribution is assumed to be stationary (as stated in Equation \ref{eq:assump1}), and (ii) the marginal system state distribution $\pi$ is also stationary. Consequently, the marginal performance distribution retains its stationarity.
Stationarity of the marginal distribution implies that the overall long-term behavior of the system, in terms of performance, is consistent. 
While the system might experience temporary fluctuations due to different states and observations, the overall probability of experiencing various levels of performance remains the same over time. 
This is reasonable for systems that are designed to operate within certain performance bounds in the long run.} 

To formulate predictability, we rely on the forecast and the marginal distributions defined above.
An event is considered unpredictable not because its distribution is entirely unknown, but because the forecast distribution does not significantly differ from the marginal distribution.
Based on that, we define predictability in the following definition.
\begin{definition} \label{def:predictability}
A system is unpredictable with lead time $L$ if the forecast and marginal distributions do not differ if
\begin{equation}
    \Pr(Z_{n+L} \mid O_{0:n} = o_{0:n}) = \Pr(Z_{n+L}).
\end{equation}
According to this definition, the loss of predictability occurs when the future state $Z_{n+L}$ is statistically independent of the observations.
Equivalently, we define the degree of predictability as the \gls{TV} distance between the forecast and marginal distributions
\begin{equation}
     D_n(L) = \norm{\Pr(Z_{n+L}\mid O_{0:n} = o_{0:n}) - \Pr(Z_{n+L}))}_{\text{TV}},
     \label{eq:predictability}
\end{equation}
where the \gls{TV} distance is a statistical metric distance defined by
\begin{align}
     \text{TV}(p,q) &:= \sup_{B \subset \mathcal{Z}} |p(B) -  q(B)|\\
     &= \frac{1}{2} \sum_{z \in \mathcal{Z}} | p(z) - q(z) |.
\end{align}
for probability mass functions $p$ and $q$.
\bluewrap{It is important to emphasize that this definition relies on the marginal distribution being stationary.}
\end{definition}

The \gls{TV} distance is a metric that ranges between zero and one and benefits from numerous properties for bounding and approximation.
Intuitively, the system is unpredictable when the \gls{TV} distance between the forecast and the marginal is very small, so the performance and observations are almost independent.
On the contrary, the system is predictable when \gls{TV} is close to 1 since the distributions are very different and that means the observations make a difference in our prediction of the performance.
This framework implies that predictability is a combined property of the dynamical system and the observations.
The observations might be irrelevant, leading to zero predictability, or the system conditions might change in a completely random manner, rendering the incorporation of the observations ineffective.

It is important to acknowledge that the predictability framework introduced in this paper may not be universally applicable to all systems. 
Specifically, it might seem counterintuitive when applied to deterministic or nearly deterministic processes that do not need side information. 
For example, a process that always takes the value 1 would be considered unpredictable according to this definition because the forecast and marginal distributions are identical, yielding a total variation distance of zero. 
However, in such a scenario, alternative measures like entropy could be more appropriate, as they would correctly identify the process as fully predictable due to its zero entropy. 
Thus, the proposed definition is particularly well-suited for systems where the primary concern is the utility of observations in reducing uncertainty.
This is especially relevant for networked systems where the decision to incorporate additional signals in performance prediction must be weighed against the cost of processing and routing them.

\begin{definition} \label{def:predictabletime}
Epsilon-predictable horizon of a system is defined by the maximum lead time that predictability remains greater than $\epsilon$:
\begin{equation}
    H_{\epsilon} = \max\{ L, \forall n : D_{n}(L) \geq \epsilon \}.
\end{equation}
\end{definition}
This metric guarantees that if we have observations up to time $n$, then we will have informative forecasts up to time $n+L$, since the forecast distribution will always have minimum distance $\epsilon$ with the marginal distribution.
Essentially, it reflects how quickly the observations lose their importance, as the system evolves quickly to its stationary state.
Therefore, the epsilon-predictable horizon is essential in analyzing the predictability of complex dynamic systems.

In addressing the problems considered in this study, we begin by assessing the predictability of system models under conditions of imperfect observations.
Specifically, we aim to determine the predictability of the systems with the following forecast distributions:
\begin{itemize}
    \item with delayed observation
    \begin{equation}
        \Pr(Z_{n+L}\mid O_n = X_{n-d}),
    \end{equation}
    \item with the observation of aggregated states
    \begin{equation}
        \Pr(Z_{n+L}\mid O_{n} = \phi(X_n)),
    \end{equation}
    \item and with partial observation in multihop cases
    \begin{equation}
        \Pr(Z_{n+L}\mid O_{n} \subset \{X_n^{(1)},\cdots,X_n^{(M)}\}).
    \end{equation}
\end{itemize}

Next, we will determine upper bounds on predictability based on $L$ for both single-hop and multi-hop system models, which can then be converted to an upper limit on $H_\epsilon$.
Finally, we will derive exact and approximation solutions for the predictability of sojourn time in basic queuing systems, specifically Geo/Geo/1/K.

We will evaluate the results in both single-hop and multihop scenarios under various configurations with imperfect observations and state aggregation.
Building on the system model and predictability definitions established in this section, we now delve into a detailed predictability analysis for Markov-modulated systems in the following section.

%% file: 4abc_predictability.tex
\section{Predictability Analysis}

To begin, we consider perfect observation of conditions, as $O_n = X_n$ for all $n$, until otherwise specified.
In the following lemma, we introduce two immediate conclusions which apply to any Markov-modulated probabilistic performance measure.
\begin{lemma}\label{lemma:mixture}
In a system where the performance $Z_n$ is related to Markov chain conditions $X_n$ via a known stationary posterior distribution $\Pr(Z_n \mid X_n)$, forecast distribution (distribution of performance $L$ time slots ahead, given the observed state $x$) can be derived as
\begin{align}
\begin{split}
    \Pr(Z_{n+L} \mid X_{n} = x) &= \sum_{y \in \mathcal{X}} P^{L}(x,y) \Pr(Z_n \mid X_n = y)
\end{split}
\label{eq:gfc}
\end{align}
Marginal distribution of performance can be obtained via
\begin{align}
\begin{split}
    \Pr(Z_{n+L}) &= \sum_{y \in \mathcal{X}} \pi(y) \Pr(Z_n \mid X_n = y).
\end{split}
\end{align}
Proof: we begin by expanding forecast definition using law of total probability as
\begin{multline}
\Pr(Z_{n+L} = z \mid X_n = x) = \\
\sum_{y \in \mathcal{X}} \Pr(Z_{n+L} = z \mid X_{n+L} = y, X_n = x)  \\ 
\Pr(X_{n+L} = y \mid X_n = x).
\end{multline}
Then, according to first-order Markov property:
\begin{multline}
\Pr(Z_{n+L} = z \mid X_n = x) = \\ 
\sum_{y \in \mathcal{X}} \Pr(Z_{n+L} = z \mid X_{n+L} = y) \\
\Pr(X_{n+L} = y \mid X_n = x).
\end{multline}
The definition of Markov chain's transition matrix lets us simplify more:
\begin{multline}
\Pr(Z_{n+L} = z \mid X_n = x) = \\ 
\sum_{y \in \mathcal{X}} \Pr(Z_{n+L} = z \mid X_{n+L} = y) P^L(x, y) = \\
\sum_{y \in \mathcal{X}} r_y(z) P^L(x, y).
\end{multline}
For the marginal distribution, again we expand it using law of total probability as
\begin{multline}
    \Pr(Z_{n+L} = z) = \sum_{y \in \mathcal{X}} \Pr(Z_{n+L} = z \mid X_{n+L}=y) \\ \Pr( X_{n+L}=y ).
\end{multline}
Moreover, we use the definition of Markov chain's stationary state probabilities to finish the proof
\begin{multline}
\Pr(Z_{n+L} = z) = \\ 
\sum_{y \in \mathcal{X}} \Pr(Z_{n+L} = z \mid X_{n+L} = y) \pi(y) = \sum_{y \in \mathcal{X}} r_y(z) \pi(y).
\end{multline}
\end{lemma}

This lemma indicates that in such a system, forecast distribution becomes a mixture model with kernels defined by the posterior distribution and weights defined by Markov chain's transient state probabilities.
As for the marginal distribution, regardless of the observations, it also becomes a mixture model with the same kernels but weights are defined by the Markov chain's stationary state probabilities

\begin{theorem}\label{theorem:pred_derivation}
Predictability of a Markov-modulated process $Z_{n}$ with Markov chain probabilities \{$P$, $\pi$\} and posterior distributions $r_y(z)$, based on the perfect observation of the state $x$ is characterized as follows:
\begin{align}
    D_n(L) &= \sup_{A \subset \mathcal{Z}} \big\lvert \sum_{z \in A}  \sum_{y \in \mathcal{X}}(P^{L}(x,y) - \pi(y) )r_{y}(z) \big\rvert,\\
    &= \frac{1}{2} \sum_{z \in \mathcal{Z}} \lvert \sum_{y \in \mathcal{X}}(P^{L}(x,y) - \pi(y) )r_{y}(z) \rvert.
\end{align}
Proof: Substituting the forecast and marginal distributions in lemma \ref{lemma:mixture} into the predictability definition \ref{def:predictability} results the two forms for predictability derivation.
\end{theorem}
This theorem provides an inaugural methodology for deriving the predictability of a general Markovian system.
It requires characterization of the conditions as a Markov model to determine $\{P, \pi\}$ and the derivation of the conditional distribution of the target metric across all conditions $r_{y}(z)$ $\forall y$ or the posterior distribution.
In data-driven scenarios, these tasks correspond to Markov model parameter estimation and conditional density estimation, respectively.

In the following subsections, we address more specific scenarios with imperfect observations, spectral analysis, multihop, and queueing systems.

\subsection{Imperfect Observations Impact}
We analyze the impact of imperfect observations on predictability in two parts:
\paragraph{Delayed Observations}
This scenario is particularly critical given a network of remote nodes involved. 
Observations must be transmitted to the forecast location, introducing potential delays where $O_n = X_{n-d}$.
\begin{lemma} \label{lemma:delay}
    The predictability of a Markov-modulated process for the lead time $L$ given observations delayed by $d$ timeslots equals the predictability of the system with perfect observations but for the lead time $L+d$.\\
    Proof: Suppose an observation arrives at the forecast point with a delay of $d$ timeslots. 
    The forecast distribution for this scenario becomes $\text{Pr}(Z_{n+L} \mid X_{n-d})$.
    Given that the Markov chain and posterior distribution are time invariant, we can substitute the current time index $n$ with $n+d$ as follows 
    \begin{equation}
        \text{Pr}(Z_{n+L} \mid X_{n-d}) = \text{Pr}(Z_{n+L+d} \mid X_n).
    \end{equation}
    Therefore, $L+d$ can be seen as the new forecast lead time while other variables in Theorem $\ref{theorem:pred_derivation}$ remain the same.
\end{lemma}
When the observation delay increases, it eventually exceeds the epsilon-predictable horizon, and the forecast converges to the marginal distribution, rendering the observation ineffective.
Therefore, the delay in the observations must be smaller than the epsilon-predictable horizon of the system to have useful predictions.

\paragraph{Aggregated States Observation}
In this scenario, the aggregation results in a new Markov chain with probabilities $\{ \bar{P}, \bar{\pi} \}$ and a new posterior distribution $\bar{r}_{a}(z)$ for any $a \in \mathcal{A}$.
Since Lemma \ref{lemma:mixture} and Theorem \ref{theorem:pred_derivation} would hold true for this new aggregated setup, we can derive predictability accordingly.
In the following lemma, we demonstrate how to construct $\bar{P}$, $\bar{\pi}$, and $\bar{r}_a(z)$.
\begin{lemma} \label{lemma:aggregation}
    Suppose the states of a Markov chain $\mathcal{X}$ are aggregated via the surjective map $\phi$ resulting a new Markov chain with state space $\mathcal{A}$ and probabilities $\{\bar{P},\bar{\pi}\}$.
    The aggregate transition matrix denoted by $\bar{P}$ is derived by 
    \begin{equation} \label{eq:aggprob}
        \bar{P}(a, b) = \frac{\sum_{x \in \phi^{-1}(a)} \sum_{y \in \phi^{-1}(b)} P(x, y) \pi(x)}{\pi(\phi^{-1}(a))}.
    \end{equation}
    The posterior distribution is obtained via
    \begin{equation} \label{eq:aggstat}
        \bar{r}_{a}(z) = \frac{\sum_{y \in \phi^{-1}(a)} \pi(y) r_{y}(z) }{\pi(\phi^{-1}(a))}.
    \end{equation}
    The aggregate stationary distribution is evident to be 
    \begin{equation}
        \bar{\pi}(b) = \pi(\phi^{-1}(b)).
    \end{equation}
    Proof: We begin with expanding the transition probability matrix using the law of total probability:
    \begin{multline}
    \bar{P}(a, b) = \Pr(\phi(X_{n+1}) = b \mid \phi(X_n) = a) = \\
    \frac{\sum_{x \in \phi^{-1}(a)} \Pr(\phi(X_{n+1}) = b \mid X_n = x) \Pr(X_n = x)}{\Pr(\phi(X_n) = a)}.
    \end{multline}
    Then, using additivity we can substitute the numerator as
    \begin{multline}
    \sum_{x \in \phi^{-1}(a)} \Pr(\phi(X_{n+1}) = b \mid X_n = x) \Pr(X_n = x) = \\
    \sum_{x \in \phi^{-1}(a)} \sum_{y \in \phi^{-1}(b)} \Pr(X_{n+1} = y \mid X_n = x) \Pr(X_n = x),
    \end{multline}
    and use Markov chain notations to achieve Equation \ref{eq:aggprob}.
    For the posterior distribution we apply simillar approach and obtain
    \begin{multline}
        \bar{r}_{a}(z) = \Pr( Z_n = z \mid \phi(X_n) = a) = \\
    \frac{\sum_{y \in \phi^{-1}(a)} \Pr(Z_n = z \mid X_n = y) \Pr(X_n = y)}{\Pr(\phi(X_n) = a)},
    \end{multline}
    where by substituting notations we get Equation \ref{eq:aggstat}.
\end{lemma}

In the next section we opt for a more intuitive and robust understanding of predictability by analyzing the inherent properties of the Markov chain with spectral analysis, rather than solely relying on the estimation/assumption of state transition probabilities.

\subsection{Markov Chain Spectral Properties Impact}

In the following lemma, we summarize the literature on spectral analysis which is applicable on the total variation distance to equilibrium in reversible Markov chains.

\begin{lemma}[\cite{Winkler2003,sousi_mixing_2020}] \label{lemma:mcupperbound}
    In a reversible Markov chain, for any starting state $x$, the total variation distance to stationary state probabilities is bounded as
    \begin{equation} \label{eq:mc-upperbound1}
         \left\| P^{L}(x, \cdot)-\pi\right\|_{\text{TV}} \leq \frac{1}{2} \left( \sum_{j=2}^{K} \lambda^{2L}_{j} f^2_j(x) \right)^{1/2}.
    \end{equation}
    Suppose $\lambda_{*} = \text{max}\{|\lambda|, \lambda \neq 1\}$, then $\xi = 1-\lambda_{*}$ denotes the \textit{spectral gap} which can simplify the bound further:
    \begin{equation} \label{eq:mc-upperbound}
        \left\| P^{L}(x, \cdot)-\pi\right\|_{\text{TV}} \leq \frac{1}{2}\lambda_{*}^L \left(\frac{1}{\pi(x)}-1\right)^{1/2}.
    \end{equation}
\end{lemma}
Spectral gap is known to provide a crucial measure on the rate at which the chain converges to its stationary distribution.
This measure quantifies the chain's mixing properties: a larger spectral gap implies faster convergence, meaning that the distribution of the chain's states will quickly approach the stationary distribution.

By employing the spectral analysis principles, we derive an upper bound on the predictability of general Markovian systems which is introduced in the following theorem.
\begin{theorem}[Spectral-Based Predictability Bound] \label{theorem:upperbound}
Predictability of a Markov-modulated process $Z_{n}$ with Markov chain probabilities \{$P$, $\pi$\} and stationary posterior distribution $r_y(z)$, based on the perfect observation of the state $x$ is upper-bounded as follows:
\begin{equation} \label{eq:pred-upperbound1}
D_n(L) \leq \frac{1}{2} \left( \sum_{j=2}^{K} \lambda^{2L}_{j} f^2_j(x) \right)^{1/2} \sqrt{2} \left( R - 1\right)^{1/2},
\end{equation}
Where
\begin{equation}
R = \sum_z \frac{\sum_{y} \pi(y)r^{2}_y(z)}{\sum_y{\pi(y)r_y(z)}}, \text{ and } 1 \leq  R \leq K.
\end{equation}
The inequality can be more simplified using the second largest eigenvalue as
\begin{equation} \label{eq:pred-upperbound2}
    D_n(L) \leq \frac{1}{2} \lambda^{L}_{*} \left(\frac{1}{\pi(x)} - 1\right)^{1/2} \sqrt{2} \left( R - 1\right)^{1/2}.
\end{equation}
Proof: \emph{see appendix \ref{app:upperbound}}.
\end{theorem}
This theorem suggests that the predictability upper bound decays geometrically, with the same rate that Markov chain converges to stationarity.
Upper bounds in Equations \ref{eq:pred-upperbound2} and \ref{eq:mc-upperbound} are different in only one term which is sensitive to the closeness of posterior distributions $r_y(z)$.
The minimum value of $R$, i.e. $1$, occurs when all conditional distributions are similar and makes the predictability zero as was expected.
On the other hand, we get the maximum, i.e. $K=|\mathcal{X}|$, when conditional distributions have minimal overlap.

\begin{remark}
    For the cases that $\frac{3}{2} < R \leq K$, a tighter upper bound on the predictability is the Markov chain's total variation distance to equilibrium. \\
    Proof: using triangle inequality and $0 \leq r_y(z) \leq 1 \quad \forall y \in \mathcal{X}, \forall z \in \mathcal{Z}$ on Theorem \ref{theorem:pred_derivation} we have
    \begin{multline}
        D_n(L) = \frac{1}{2} \sum_{z \in \mathcal{Z}} \bigg| \sum_{y \in \mathcal{X}}  \left( P^L(x, y) - \pi(y) \right) r_y(z) \bigg| \\
        \leq \frac{1}{2} \sum_{z \in \mathcal{Z}} \sum_{y \in \mathcal{X}} \left( \left| P^L(x, y) - \pi(y) \right| r_y(z) \right) \\
        = \frac{1}{2} \sum_{y \in \mathcal{X}} \left| P^L(x, y) - \pi(y) \right| \sum_{z \in \mathcal{Z}} r_y(z) \\
        = \frac{1}{2} \sum_{y \in \mathcal{X}} \left| P^L(x, y) - \pi(y) \right| = \frac{1}{2} \|P^L(x, \cdot) - \pi\|_{\text{TV}}.
    \end{multline}
\end{remark}
Theorem \ref{theorem:upperbound} introduces a novel approach to analyzing observable Markov models that modulate a secondary process through a known, stationary posterior distribution. 
Unlike traditional Markov chain mixing time analyses that focus on the total variation distance between stationary and transient state probabilities, our work provides an upper bound on the total variation distance between the secondary processes modulated by these different state probabilities. 
This advancement offers a new perspective on the convergence behavior of complex, multi-layered stochastic systems.

The subsequent section explores techniques for analyzing predictability in multi-hop systems.

\subsection{Predictability in Multi-Hop Systems}
End-to-end performance predictability analysis in a multi-hop system as shown in Figure \ref{fig:sysmodel} requires an assumption on the performance measure and how it is obtained in total.
In scenarios involving multiple hops, we consider $Z_n$ to represent the end-to-end delay, which is determined as $Z_{n} = \sum_{m=1}^{M} Z_{n}^{(m)}$.
Due to the summation, total delay distribution will be the convolution of each hop's delay distribution both for forecasts and marginals as
\begin{multline} \label{eq:multihop1}
    \Pr(Z_{n+L}\mid X_{n} = x) = \\
    \Pr(Z^{(1)}_{n+L}\mid X^{(1)}_{n} = x^{(1)}) \ast \Pr(Z^{(2)}_{n+L}\mid X^{(2)}_{n} = x^{(2)}) \ast \\ \dots\ast \Pr(Z^{(M)}_{n+L}\mid X^{(M)}_{n} = x^{(M)}),
\end{multline}
\begin{multline} \label{eq:multihop2}
    \Pr(Z_{n+L}) = \\
    \Pr(Z^{(1)}_{n+L}) \ast \Pr(Z^{(2)}_{n+L}) \ast \dots\ast \Pr(Z^{(M)}_{n+L}),
\end{multline}
where for each hop, marginal and forecast distributions can be derived via Lemma \ref{lemma:mixture}.

Deriving predictability using the above method becomes less tractable as system complexity increases.
In this context, identifying approximation or bounding techniques applicable to the problem is helpful.
Below we propose a key lemma regarding the predictability in tandem systems where the end-to-end delay is subject to forecast.
\begin{lemma}[Subadditivity of Predictability] \label{lemma:multihop}
    The predictability of a tandem multi-hop system, assuming independent hops, is upper-bounded by the sum of the predictability of each hop as follows:
    \begin{equation} \label{eq:multihopproposition}
        D_n(L) \leq \sum_{m=1}^{M} D^{(m)}_n(L).
    \end{equation}
    Proof: For independent random variables $F^1$, $F^2$, $M^1$, and $M^2$ and their distributions, the total variation distance is subadditive:
    \begin{equation}
        \norm{(F^1+F^2) - (M^1+M^2)}_{\text{TV}} \leq \norm{F^1 - M^1}_{\text{TV}} + \norm{F^2 - M^2}_{\text{TV}}.
    \end{equation}
    The proof completes by supposing $F^i$ and $M^i$ denote forecast and marginal distributions of hop $i$.
\end{lemma}
This lemma proves beneficial especially when examining the predictability of a multihop system with incomplete observability i.e. $O_n \subset X_n$.
In deriving the bound by adding the predictability across all hops, if the state in one or more hops is unobserved, their predictability drops to zero and does not affect the summation in Equation $\ref{eq:multihopproposition}$.
We will investigate this in the numerics section in more detail.

%% file: 4d_predictability.tex
Next, we opt to apply the predictability analysis to Geo/Geo/1 queue, which is the discrete time analogue of the M/M/1 queue.

\subsection{Geo/Geo/1/K Queue Predictability} \label{sec:mm1-perfect}
\bluewraptwo{In this subsection, we apply the previously described predictability concepts and methods to the Geo/Geo/1/K queue model, where \(K\) denotes the maximum queue length or buffer capacity. We focus on the sojourn time (or packet delay) \(Z_n\) as the primary performance metric for our forecasts. Finally, we offer an investigation on how predictability definitions extend to the case in which \(Z_n\) represents the queue’s blocking to model packet loss ratio instead.}
The transition matrix in this queue is given by
\begin{multline} \label{eq:mm1prob}
\Pr(X_n = j \mid X_{n-1} = i) = \\
\begin{cases}
\alpha\bar{\mu}, & \text { if } j=i+1,0 \leq i \leq K-1 \\ 
\mu\bar{\alpha}, & \text { if } j=i-1,1 \leq i \leq K \\ 
1-\alpha\bar{\mu}, & \text { if } j=i=0 \\ 
\alpha \mu+\bar{\alpha}\bar{\mu}, & \text { if } j=i, 1 \leq i \leq K-1 \\ 
1-\mu\bar{\alpha}, & \text { if } j=i=K \\ 
0, & \text { otherwise }
\end{cases}
\end{multline}
and the corresponding Markov chain is depicted in Figure \ref{fig:mm1}.
\begin{figure}[!ht]
\centering
\begin{tikzpicture}[->, >=stealth', auto, semithick, node distance=2cm]
    \tikzstyle{every state}=[fill=white,draw=black,thick,text=black,scale=1, minimum size=1cm, text width=5mm, align=center]
    
  \node[state] (0) {$0$};
  \node[state] (1) [right of=0] {$1$};
  \node[state] (dots) [right of=1] {$\dots$};
  \node[state] (c) [right of=dots] {\scalebox{0.7}{$K$}};

  \path[every node/.style={font=\sffamily\small}]
    (0) edge[loop above, above] node{$1-\alpha\bar{\mu}$}	(0)
    (0) edge [bend left] node[above]{$\alpha\bar{\mu}$} (1)
    
    (1) edge[loop above, above] node{$\alpha \mu+\bar{\alpha}\bar{\mu}$}	(1)
    (1) edge [bend left] node[above]{$\alpha\bar{\mu}$} (dots)
    (1) edge [bend left] node[below]{$\mu\bar{\alpha}$} (0)
    

    (dots) edge[loop above, above] node{$\alpha \mu+\bar{\alpha}\bar{\mu}$}	(dots)
    (dots) edge [bend left] node[above]{$\alpha\bar{\mu}$} (c)
    (dots) edge [bend left] node[below]{$\mu\bar{\alpha}$} (1)

    (c) edge[loop above, above] node{$1-\mu\bar{\alpha}$}	(c)
    (c) edge [bend left] node[below]{$\mu\bar{\alpha}$} (dots);
\end{tikzpicture}
\caption{Geo/Geo/1/K Markov chain states and transition probabilities}
\label{fig:mm1}
\end{figure}

In the following theorem we describe an exact derivation of predictability for this queuing model based on the observation of queue length and Theorem $\ref{theorem:pred_derivation}$.
\begin{theorem}
\label{theorem:geogeo1pred}
    The predictability of delay in a Geo/Geo/1/K queue given the observation of the system size denoted by $x$ is described as follows:
    \begin{multline}\label{eq:mm1pred}
        D_n(L) = \\
        \frac{1}{2}\sum_{z} \biggl| \sum_{y=1}^{K} \frac{2}{K+1}\beta^{\frac{y-x}{2}} \sum_{k=1}^{K} \frac{\gamma_k^L}{\left[1-2 \sqrt{\beta} \cos \left(\frac{k \pi}{K+1}\right)+\beta\right]} \\
        \left[\sin \left(\frac{x k \pi}{K+1}\right)-\sqrt{\beta} \sin \left(\frac{(x+1) k \pi}{K+1}\right)\right] \\
        \left[\sin \left(\frac{y k \pi}{K+1}\right)-\sqrt{\beta} \sin \left(\frac{(y+1) k \pi}{K+1}\right)\right] \text{NB}(z; y,\mu) \biggr|,
    \end{multline}
    where 
    \begin{equation*}
        \gamma_{k} = \alpha \mu+\bar{\alpha}\bar{\mu}+ 2 \sqrt{\alpha \mu\bar{\alpha}\bar{\mu}} \cos \left(\frac{k \pi}{K+1}\right) \text{ and } \beta = \frac{\alpha\bar{\mu}}{\mu\bar{\alpha}}.
    \end{equation*}
    Proof: As Theorem \ref{theorem:pred_derivation} indicates, to analyze the predictability of this system with perfect observations, we require the derivation of 3 components: state probabilities for a lead time $L$, given an observed state $x$, stationary probabilities of states $\pi(\cdot)$, and the posterior distribution $\Pr(Z_n \mid X_n = x)$.
    The posterior distribution in this case is the sojourn time distribution conditioned on the queue size.
    For a Geo/Geo/1 queue with service probability $\mu$ and $x$ packets in the system, the sojourn time distribution is known as
    \begin{equation}
        \label{eq:mm1conds}
        \Pr(Z_n \mid X_n = x) \sim \text{NB}(x+1,\mu),
    \end{equation}
    where $\text{NB}$ denotes the negative binomial density function.
    As for the transient state probabilities $L$ timeslots ahead, they could be derived by raising the transition matrix to $L$ i.e. $P^{L}(x,\cdot)$ or using its another form derived in \cite{kim2008transient} given by:
    \begin{multline}
    \label{eq:mm1tans}
        P^L(i,j) = \\
        = \frac{1-\beta}{1-\beta^{K+1}}\beta^j+\frac{2}{K+1}\beta^{\frac{j-i}{2}} \sum_{k=1}^{K} \frac{\gamma_k^L}{\left[1-2 \sqrt{\beta} \cos \left(\frac{k \pi}{K+1}\right)+\beta\right]} \\
        \left[\sin \left(\frac{i k \pi}{K+1}\right)-\sqrt{\beta} \sin \left(\frac{(i+1) k \pi}{K+1}\right)\right] \\
        \left[\sin \left(\frac{j k \pi}{K+1}\right)-\sqrt{\beta} \sin \left(\frac{(j+1) k \pi}{K+1}\right)\right].
    \end{multline}
    When the queue is stable, i.e. $\mu > \alpha$, the stationary probabilities are known as
    \begin{equation}
    \label{eq:mm1stat}
    \pi(y)
    = \frac{1-\beta}{1-\beta^{K+1}}\beta^y.
    \end{equation}
    Then, we derive predictability by replacing Equations \ref{eq:mm1conds}, \ref{eq:mm1tans}, and \ref{eq:mm1stat} in Theorem \ref{theorem:pred_derivation} to complete the proof.
\end{theorem}

We propose an alternative approach to the aforementioned method for two reasons. 
Firstly, because of the possible computational demands, and secondly, because of the difficulty in determining the relationship between model parameters and predictability.
Hence, we present this approximation in the next proposition.
\begin{proposition}\label{proposition:geogeo1approx}
    The predictability of the sojourn time of a Geo/Geo/1/K queue with a very large $K$ for $L$ time slots in future based on the observation of the queue size $x$ is approximated as
    \begin{multline}
        D_n(L) \approx \frac{\beta^{\frac{1-x}{2}}}{\pi}\left(1-\sqrt{\beta}\right)^2\left(\alpha\mu + \bar{\alpha} \bar{\mu} + 2\sqrt{\alpha\mu\bar{\alpha}\bar{\mu}}\right)^{L} \\ \int_{0}^{\pi} \frac{\sin{r}\sin{x r}}{(1-2\sqrt{\beta}\cos{r}+\beta)^2} e^{-L \kappa r^2} dr,
    \end{multline}
    where
    \begin{equation}
        \kappa = \frac{\sqrt{\alpha\mu\bar{\alpha}\bar{\mu}}}{\alpha\mu + \bar{\alpha} \bar{\mu}+2\sqrt{\alpha\mu\bar{\alpha}\bar{\mu}}}.
    \end{equation}
    Proof: \emph{see appendix \ref{app:geogeo1approx}}.
\end{proposition}
Compared to Theorem \ref{theorem:geogeo1pred}, using this approximation we can focus only on the terms that are critical in terms of influence on predictability.
For instance, the term outside the integral which is raised to $L$.
If it is reduced, as $L$ grows, predictability diminishes more rapidly.
This implication necessitates understanding when the approximation is valid and incorporating that understanding into the problem at hand.
For example, as indicated in the proof, the approximation assumes large value for $K$.
We will evaluate this in more detail in the numeric section.

\bluewraptwo{
Finally, as noted previously, a Geo/Geo/1/K queue is a lossy system that may reject or drop arriving packets if the queue is full. We demonstrate that our proposed predictability framework can capture the packet loss ratio in this queueing scenario. Specifically, let \(Z_n\) be an indicator variable describing whether an arriving packet at slot \(n\) encounters a full queue, with \(Z_n \in \{0,1\}\). Hence, \(\Pr(Z_n = 1)\) is the packet loss ratio at slot \(n\), which we take as our performance metric in place of delay. In the following proposition, we provide an analytical derivation of the packet loss ratio's predictability for a Geo/Geo/1/K queue.
\begin{proposition}\label{proposition:geogeo1loss} 
    The predictability of blocking in a Geo/Geo/1/K queue for $L$ time slots in the future based on the observation of the queue size $x$ is given by
    \begin{multline}
        D_n(L) = \biggr| \frac{2}{K+1}\beta^{\frac{K-x}{2}} \sum_{k=1}^{K} \frac{\gamma_k^L}{\left[1-2 \sqrt{\beta} \cos \left(\frac{k \pi}{K+1}\right)+\beta\right]} \\
        \left[\sin \left(\frac{x k \pi}{K+1}\right)-\sqrt{\beta} \sin \left(\frac{(x+1) k \pi}{K+1}\right)\right]
        \sin \left(\frac{K k \pi}{K+1}\right) \biggr|.
    \end{multline}
    \end{proposition}
    \begin{IEEEproof}
     Since \(Z_n\) takes values in \(\{0,1\}\), the definition of predictability reduces to
    \begin{equation}
    D_n(L)=\bigl| \Pr(Z_n = 1 \mid X_n = x)- \Pr(Z_n = 1)\bigr|.
    \end{equation}
    To determine the blocking probability, we apply the discrete‐time equivalent of PASTA for Bernoulli arrivals, which identifies the probability of blocking with the probability of the queue being in state \(K\). Thus, when predicting \(L\) steps ahead given \(X_n = x\), the blocking probability is 
    \begin{equation}
    \Pr(Z_n = 1 \,\mid\, X_n = x) = P^L(x, K),
    \end{equation}
    and under steady‐state conditions, it is 
    \begin{equation}
    \Pr(Z_n = 1) = \pi(K).
    \end{equation}
    Substituting these from Equations \eqref{eq:mm1tans} and \eqref{eq:mm1stat} into the predictability definition completes the proof.
\end{IEEEproof}
This proposition offers an analytic expression that quantifies how knowing the current queue size \(x\) alters the predictability of near-future blocking in a Geo/Geo/1/K queue, compared to the steady-state behavior. The result underscores a non-trivial interplay among the buffer size \(K\), the forecast horizon \(L\), and the system state \(x\). Even a relatively large \(K\) can exhibit marked short-term predictability of blocking if the queue is already near capacity, whereas for a smaller \(K\), small shifts in queue length can sharply change the likelihood of blocking. As \(L\) increases, the system forgets its starting state due to Markov mixing, causing predictability to converge to zero. We will analyze this further in Section~\ref{sec:numerical}.
} 


%% file: 5ab_numerics.tex
\section{Numerical Results}
\label{sec:numerical}
This section is dedicated to evaluating the proposed methods and examining the predictability in abstract systems in 3 subsections\footnote{MATLAB code and results are accessible at https://github.com/samiemostafavi/predictability}. 
First, we evaluate the connected vehicular scenario.
Then, spectral-based upperbound and approximations on predictability of the Geo/Geo/1/K queue are analyzed.
The last subsection is devoted to analyzing how dynamics of the queue affect its predictability and in the end multi-hop queues are analyzed when their observability is limited.

\bluewrap{Throughout the section, our focus is on the relationship between predictability and lead time $L$.  
In our plots, lead time is consistently represented as the primary independent variable on the x-axis.  
A fundamental observation across all scenarios is that predictability decreases as the lead time $L$ increases. 
This reflects the inherent difficulty in predicting the future behavior of a system as we look further ahead.  
The rate of this decrease, which is often approximately exponential, is influenced by various system parameters, such as the maximum queue size $K$, the level of state aggregation, and the presence of bottlenecks in multi-hop systems. 
We will explore these influences in detail in the following subsections.}

\subsection{\bluewrap{Predictability Under the Random Walk Model}}

\begin{figure}
    \centering
    \includegraphics[width=7.5cm]{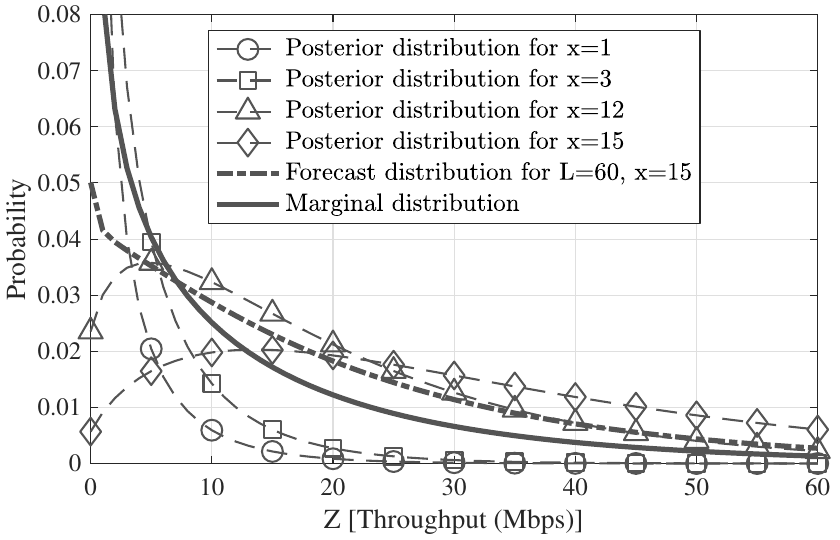}
    \caption{\bluewrap{Posterior, marginal, and forecast distributions for a cellular connection’s downlink throughput, with \gls{CQI} evolving under a random walk model with $p=0.6$ (mobile).}}
    \label{fig:condmarg}
\end{figure}

\bluewrap{As a starting point, we employ a lazy random walk to capture the dynamics of vehicular connections with varying CQI in a cellular setting.
In this scenario, the vehicle’s connection state—represented by \(X_n\) for the \gls{CQI}—follows a lazy random walk on a finite, one-dimensional state space of size \(K\), with reflecting boundaries. Specifically, with probability \(p\) the state remains unchanged, and with probability \(\tfrac{1-p}{2}\) it transitions to each of the two adjacent states (reflecting at the boundaries). This random walk has a uniform stationary distribution, \(\pi(y) = \tfrac{1}{K}\) for all \(y\).

Each \gls{CQI} state \(X_n\) ($K = 15$) modulates the network’s throughput, which we assume follows a Negative Binomial distribution \(\mathrm{NB}(z; r, q)\). We set its parameters as
\[
r = 0.105 \cdot X_n + 0.104,
\quad
q = -0.006 \cdot X_n + 0.135.
\]
These parameter values are derived from a real-world dataset in which LTE connection's downlink throughput (in Mbps) and CQI were sampled at one-second intervals for both mobile (train, bus, car) and non-mobile scenarios in Cork city, Ireland \cite{Raca2018}. See Appendix~\ref{app:randomwalk} for further details on how these parameters were derived.

From our empirical analysis, we observe that in vehicular connections the random walk remains in the same state with probability \(p = 0.6\), whereas in near-static scenarios \(p = 0.9\). This modeling approach thus links the random-walk-based CQI evolution directly to throughput distribution in mobile environments.
Figure \ref{fig:condmarg} illustrates the resulting posterior distributions given an observed connection \gls{CQI}, along with the marginal distribution derived using Lemma \ref{lemma:mixture}, and a forecast distribution for 60 seconds after when \gls{CQI} 15 was observed given the random walk model.
} 
\begin{figure}
	\centering
        \begin{subfigure}{1\linewidth}
            \centering
		\includegraphics[width=7.5cm]{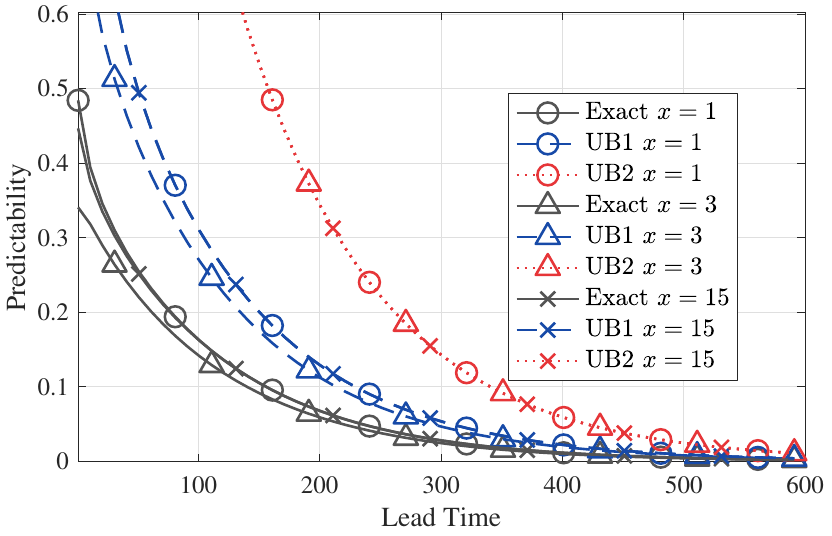}
		\caption{Mobile vehicular connection with $p=0.6$}
            \label{fig:rwub-a}
            \vspace{0.25cm}
	\end{subfigure}
	\begin{subfigure}{1\linewidth}
            \centering
		\includegraphics[width=7.5cm]{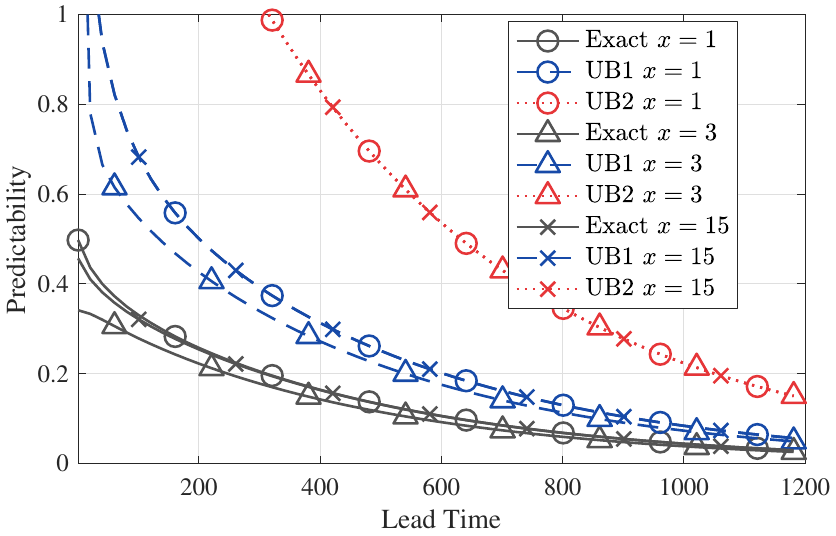}
		\caption{Static connection with $p=0.9$}
            \label{fig:rwub-b}
	\end{subfigure}
	\caption{\bluewrap{Predictability of a cellular connection’s downlink throughput, with CQI evolving under a random-walk model. The solid lines depict exact values at various lead times, while dotted and dashed lines represent two spectral upper bounds (1 and 2), respectively.}}
	\label{fig:rwub}
\end{figure}

\bluewrap{Figure \ref{fig:rwub} shows the exact predictability calculated using Theorem \ref{theorem:pred_derivation}, along with two upper bounds described in Theorem \ref{theorem:upperbound}: the full spectral-based upper bound in Equation \ref{eq:mc-upperbound1} illustrated with dashed lines (UB1) and the spectral gap-based bound in Equation \ref{eq:mc-upperbound} (UB2) illustrated with dotted lines.
Several key observations can be made from these results:
\begin{enumerate}
    \item Decay patterns: All predictability curves show a decay as the lead time increases, reflecting the gradual loss of predictive power as we look further into the future. However, predictability remains higher for a longer duration in static connections (Figure \ref{fig:rwub-b}) because the measured \gls{CQI} is more likely to stay constant, unlike the more variable behavior observed in mobile scenarios (Figure \ref{fig:rwub-a}).
    \item State-dependent predictability: The first and last states ($x=1$ and $x=15$) offer higher predictability compared to middle states (e.g., $x=3$). 
    This is because their corresponding conditional distributions differ more significantly from the marginal distribution.
    \item Upper bound tightness: For smaller lead times $L$, the upper bounds show a greater distance from the true predictability values. 
    In general, we observe that the upper bounds are tighter for predictability curves that decay more rapidly.
    \item Spectral gap bound limitations: The spectral gap-based bounds do not differentiate between observed states $x$, as seen in Equation \ref{eq:pred-upperbound2}, and are typically more relaxed due to the additional simplifications were applied to them.
    This is due to the equiprobable stationary state probabilities of the Markov chain in this case.
\end{enumerate}

These findings highlight the role of Markovian dynamics in shaping the predictability of network throughput, particularly in mobile environments where the state persistence probability influences forecasting accuracy. In the next subsection, we extend this analysis to queuing systems.
} 

\subsection{Approximation and Upper Bound Validation}

\begin{figure}
	\centering
        \begin{subfigure}{1\linewidth}
            \centering
		\includegraphics[width=7.5cm]{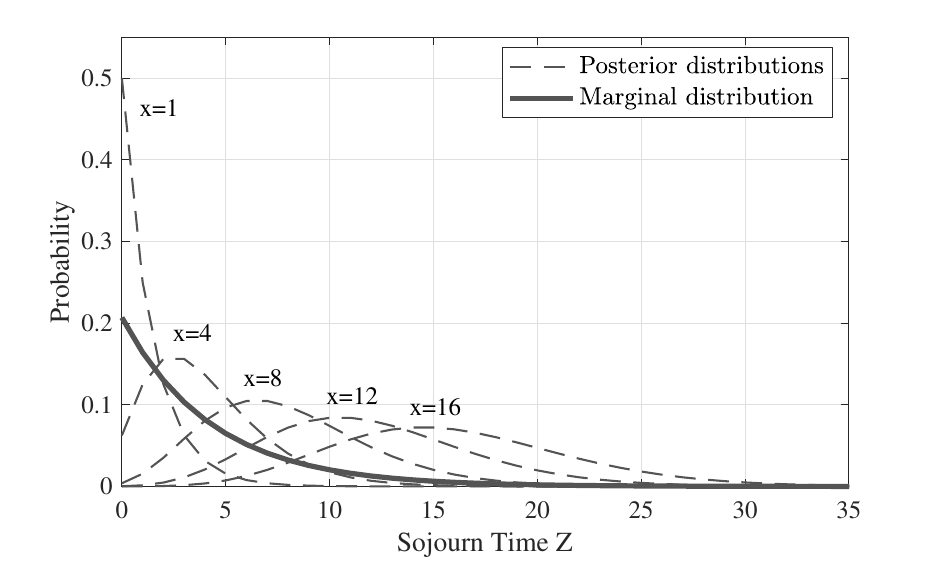}
            \vspace{0.25cm}
	\end{subfigure}
	\begin{subfigure}{1\linewidth}
            \centering
		\includegraphics[width=7.5cm]{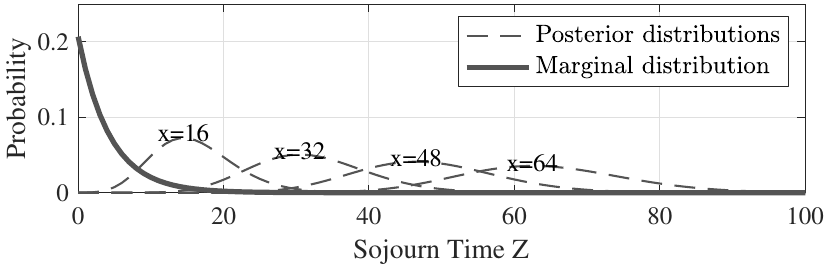}
	\end{subfigure}
	\caption{Posterior and marginal distributions of a Geo/Geo/1/K queue with \(\mu=0.5\), \(\rho=0.85\), and \(K = 128\). 
    }
    \label{fig:mm1conds}
\end{figure}

Next, we shift our focus to the queuing system and analyze its predictability.
To assess the effectiveness of our proposed approximation method in Proposition \ref{proposition:geogeo1approx} and the upper bound in Theorem \ref{theorem:upperbound}, we analyze a Geo/Geo/1/K queue with default parameters $\mu=0.5$, $\rho=0.85$, and $K=128$.
Figure \ref{fig:mm1conds} illustrates the posterior and marginal delay distributions given various observed queue lengths for this queue.

\begin{figure}
	\centering
        \begin{subfigure}{1\linewidth}
            \centering
		\includegraphics[width=7.5cm]{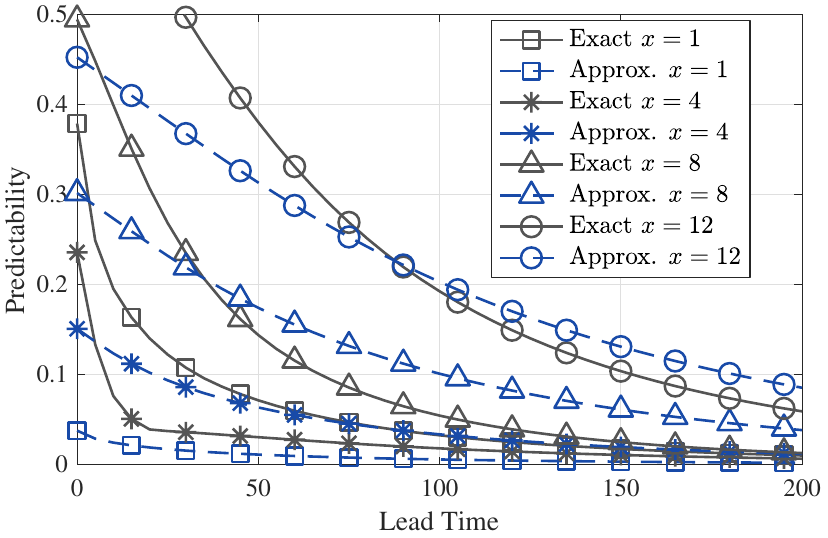}
            \vspace{0.25cm}
	\end{subfigure}
	\begin{subfigure}{1\linewidth}
            \centering
		\includegraphics[width=7.5cm]{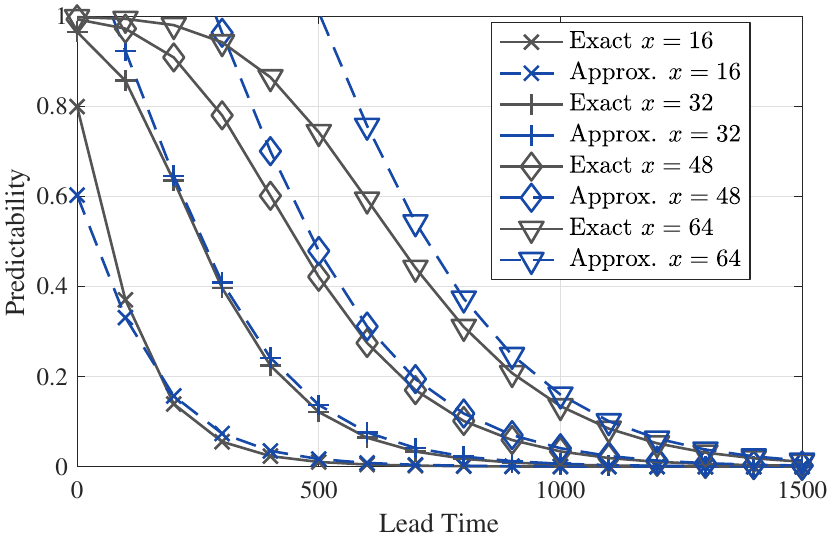}
	\end{subfigure}
	\caption{Predictability evaluation for different observed states of a Geo/Geo/1/K system with $\mu=0.5$, $\rho = 0.85$, and $K = 128$.
    The approximation method from Proposition \ref{proposition:geogeo1approx} (dashed lines) is compared against the exact predictability derivation via Theorem \ref{theorem:pred_derivation} (solid lines).}
	\label{fig:mm1approxeval}
\end{figure}

In Figure \ref{fig:mm1approxeval}, we validate the approximation method by comparing it to the exact predictability derivation presented in Theorem \ref{theorem:pred_derivation}.
\bluewrap{From the exact calculations we observe that predictability decreases as the lead time $L$ increases.}
Regarding the approximation, the main finding here is that its accuracy is significantly influenced by the disparity between the observed state's forecast distribution and the marginal distribution.
As illustrated in Figure \ref{fig:mm1conds}, states that are closer to the marginal distribution display less accurate approximations in Figure \ref{fig:mm1approxeval}, whereas states further from the marginal distribution (those above 16) demonstrate more accurate approximations.

\begin{figure}
	\centering
        \begin{subfigure}{1\linewidth}
            \centering
		\includegraphics[width=7.5cm]{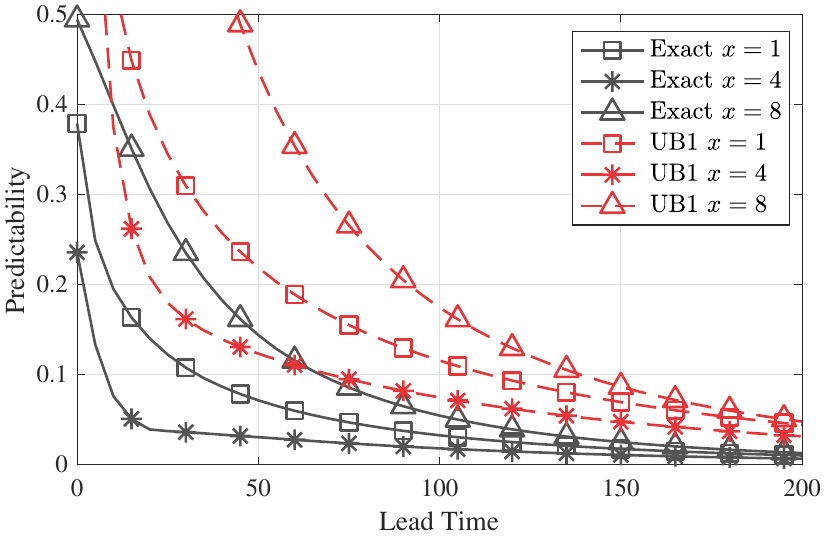}
            \vspace{0.25cm}
	\end{subfigure}
	\begin{subfigure}{1\linewidth}
            \centering
		\includegraphics[width=7.5cm]{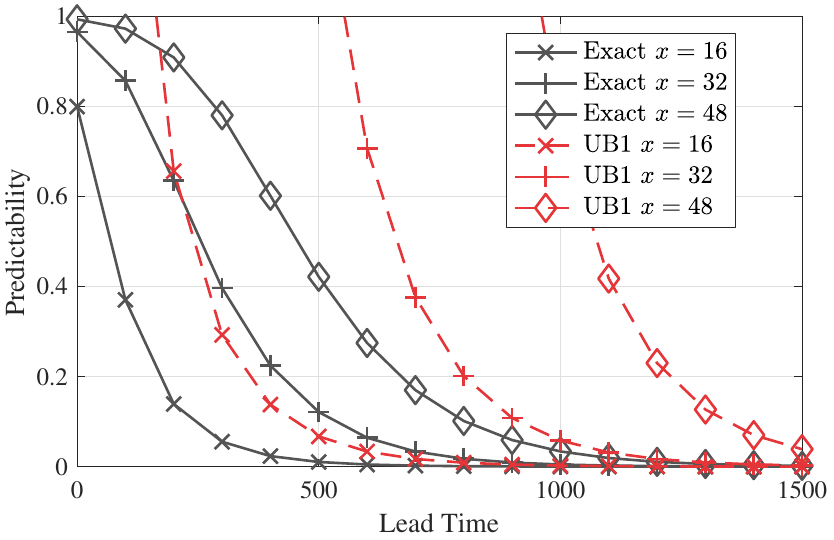}
	\end{subfigure}
	\caption{Predictability evaluation for different observed states of a Geo/Geo/1/K system with $\mu=0.5$, $\rho = 0.85$, and $K = 128$.
    The full spectral-based upper bound from Theorem \ref{theorem:upperbound} (dashed lines) is compared against the exact predictability derivation via Theorem \ref{theorem:geogeo1pred} (solid lines).}
	\label{fig:mm1ubeval}
\end{figure}

Figure \ref{fig:mm1ubeval}, shows the same setup as Figure \ref{fig:mm1approxeval}, but with the full spectral-based upper bound in Theorem \ref{theorem:upperbound} (UB1).
It is evident from the figure that unlike the approximation method, the upper bound is tighter for lower queue sizes.
As the queue size grows, the distance between the exact curve and the bound increases.
Due to poor performance, the spectral-gap based upper bound (UB2) is not involved in this analysis.

\bluewrap{This widening gap highlights a limitation of Theorem \ref{theorem:upperbound} in this specific scenario.
The theorem provides a general upper bound based on the spectral properties of the transition matrix.
However, this bound, while mathematically valid, is derived under worst-case assumptions regarding convergence to the stationary distribution. For larger queue sizes, the sojourn time predictability is primarily determined by transitions within the set of higher queue length states. A large queue tends to stay large for a while, due to the traffic intensity (\(\rho = 0.85\)). The general spectral bound, which considers all possible transitions, becomes less representative of these specific, dominant dynamics. It overestimates the influence of transitions that are less relevant when starting from a large queue, leading to a looser bound. In contrast, for smaller queue sizes, a wider range of transitions are relevant to the short-term and medium-term behavior, and the spectral bound provides a tighter approximation.}

\begin{figure}[!ht] 
  \begin{center}
    \includegraphics[width=7.5cm]{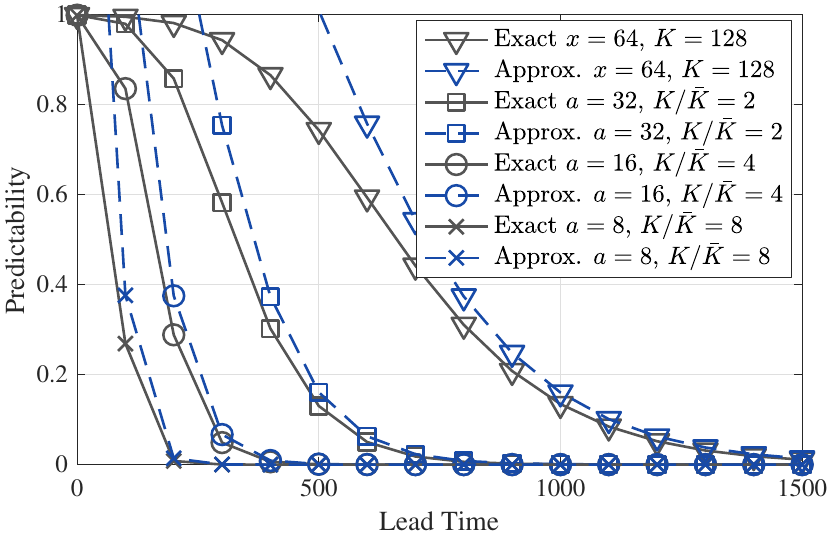}
  \end{center}
  \caption{Predictability of a Geo/Geo/1/K queue, under different state aggregation levels and $\rho=0.85$, $\mu=0.5$. Solid lines illustrate numerical derivation and dashed lines the approximation using Proposition \ref{proposition:geogeo1approx}.}
  \label{fig:mm1agg}
\end{figure}

In the next analysis, we apply state aggregation on the same queue.
Figure \ref{fig:mm1agg} illustrates four levels of sequential and even state aggregations: no aggregation, 2-state aggregation, 4-state aggregation, and 8-state aggregation.
The results show higher levels of aggregation lead to a less predictable system and more rapid decline of predictability.
The evaluation is illustrated as dashed lines in Figure \ref{fig:mm1agg} where the approximation closely follow the numerical derivation.
However, it is important to note that the current approach involves an even aggregation of states.
Future research could investigate uneven state aggregation to identify potential optimization strategies that might better balance the trade-off between observations' overhead and predictability.

%% file: 5cd_numerics.tex
\subsection{Geo/Geo/1/K Dynamics and Predictability}

To investigate the predictability in systems with Geo/Geo/1/K queues further, next, we examine the impact of the observed state and utilization factor ($\rho$).
In the following analysis, we obtain predictability using Theorem \ref{theorem:geogeo1pred} and Proposition \ref{proposition:geogeo1approx}.
\bluewrap{Furthermore, to benchmark our approach against a standard information-theoretic measure, we also quantify predictability using cross-entropy.
We define the cross-entropy predictability by:
\begin{equation}
    D^{\text{CE}}_n(L) =  - \sum_{z}  \Pr(Z_{n+L} )  \log \left( \frac{\Pr(Z_{n+L} \mid O_{0:n} = o_{0:n})}{\Pr(Z_{n+L})} \right).
    \label{eq:ce_predictability}
\end{equation}
} 

\begin{figure}
	\centering
	\begin{subfigure}{1\linewidth}
            \centering
		\includegraphics[width=7.5cm]{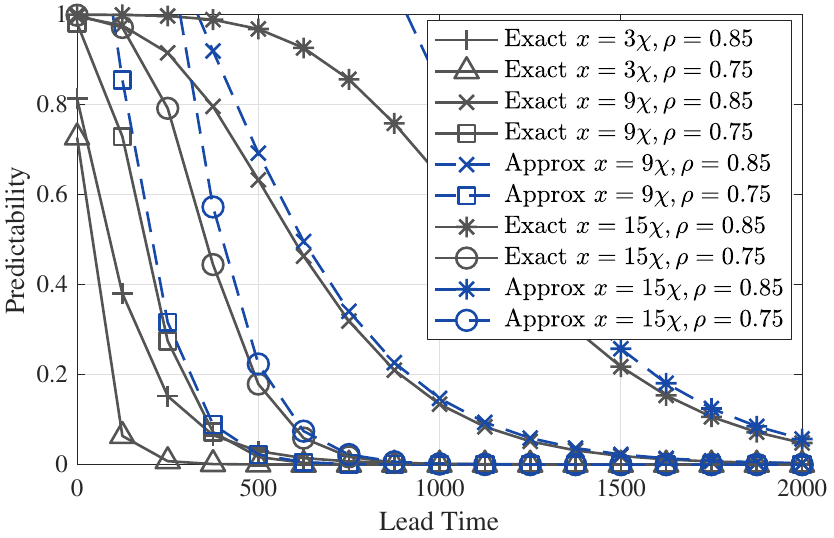}
		\caption{\Gls{TV} distance predictability}
            \label{fig:mm1preda}
	\end{subfigure}
        \begin{subfigure}{1\linewidth}
            \centering
		\includegraphics[width=7.5cm]{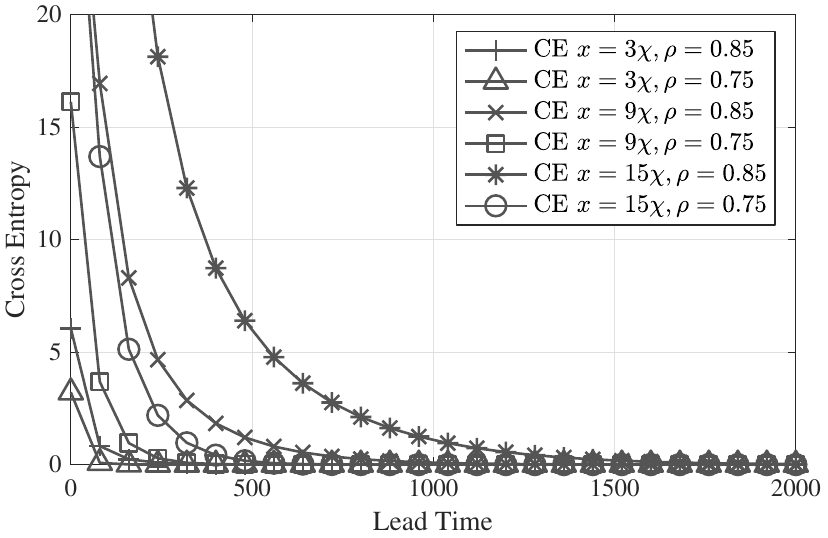}
		\caption{\bluewrap{Cross entropy (CE) predictability}}
            \label{fig:mm1predb}
	\end{subfigure}
	\caption{Predictability of a single Geo/Geo/1/K queue, under different configurations. Solid lines illustrate exact numerical derivation and dashed lines the approximation in Proposition \ref{proposition:geogeo1approx}.}
	\label{fig:mm1pred}
\end{figure}

Figure \ref{fig:mm1pred} illustrates the predictability of the Geo/Geo/1/K system, quantified using two distinct measures: \bluewrap{\gls{TV} distance (Sub-figure \ref{fig:mm1preda}) and cross-entropy predictability (Sub-figure \ref{fig:mm1predb}). }
The results are presented for parameters $\mu = 0.4$, $K = 100$ and the observed state $x$ is either 3, 9, or 15 times greater than the expected state denoted by $\chi$.
Focusing on the \gls{TV}-based predictability, it is evident from the figure that the sojourn time is more predictable when there is a significant backlog or when the queue length is substantially different from the expected queue length. 
This implies that the queue remains predictable for a certain duration into the future, and we have sufficient information to predict the delay distribution for a period of time before it converges to the stationary delay distribution.
In real systems, especially those operating in high utilization regimes where efficiency is critical, this analysis highlights the importance of monitoring the queue.
High utilization often corresponds to more significant backlog situations, where predictability is maintained for longer durations.

\begin{figure}[!ht] 
  \begin{center}
    \includegraphics[width=7.5cm]{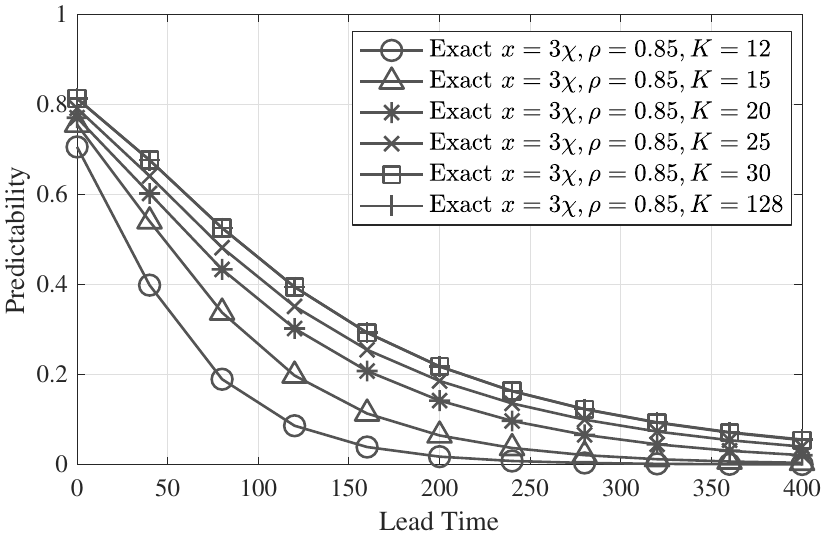}
  \end{center}
  \caption{\bluewrap{Predictability of a single Geo/Geo/1/K queue under different maximum queue sizes, derived using exact numerical calculations.}}
  \label{fig:mm1kcomp}
\end{figure}

\bluewrap{ Comparing the \gls{TV} distance (Figure \ref{fig:mm1preda}) and cross-entropy predictability (Figure \ref{fig:mm1predb}), reveals key differences arising from their mathematical definitions.  The \gls{TV} distance, being a true metric, is bounded between 0 and 1. This provides a standardized scale for assessing the degree of predictability, clearly indicating whether, and to what extent, the system's future state is influenced by its initial condition. Cross-entropy, conversely, is unbounded and can reach arbitrarily large values as observed in Figure \ref{fig:mm1predb}. This sensitivity of cross-entropy to discrepancies in low-probability regions contrasts with the more moderate response of the bounded \gls{TV} distance.  Furthermore, the established connections between \gls{TV} distance and Markov chain mixing times offer advantages in terms of analytical tractability, as demonstrated by the derivation of Theorem \ref{theorem:upperbound} in this work.

Figure \ref{fig:mm1kcomp} shows the predictability of delay in a Geo/Geo/1/K queueing system as a function of different maximum queue sizes (K), with fixed arrival and service parameters ($x = 3\chi$, $\rho = 0.85$).
The figure shows that larger queue sizes (higher $K$) result in significantly higher predictability. 
This demonstrates that systems with more buffering capacity retain information about future delays for longer, making them more predictable.
When $K$ is small, the queue has a restricted capacity for holding customers. As the offered load increases while $K$ remains limited, new arrivals are more frequently blocked or lost. This leads to a lower average queue size, as the system cannot accumulate long queues and the system reaches its maximum capacity quickly, and past information becomes less relevant.
The curves for larger $K$ values show higher predictability since the current queue size provides more information about the future system state. A longer queue now is a stronger indicator of longer delays in the future, and that information persists for a longer lead time.
However, we observe a decreasing marginal gain as $K$ increases.}

\bluewraptwo{
Building on the analysis of maximum queue size \(K\), we now focus on 
Proposition~\ref{proposition:geogeo1loss}, where performance is packet loss due to blockage rather than sojourn time. 
Figure~\ref{fig:mm1loss} shows numerical results for a Geo/Geo/1/\(K\) queue under various buffer sizes and observed queue lengths. 
When predicting near-future packet drops, an observation that the queue is nearly full (dashed lines) leads to higher predictability compared to an empty queue (solid lines). 
Further, when the queue is empty at present, having a smaller buffer increases predictability at short lead times: fewer packets are needed to overflow the system, so either “no blocking” or “rapid filling” scenarios become more pronounced and predictable. Conversely, if the queue is full and \(K\) is large, there is a strong likelihood that blocking persists in the short run, making it comparatively easier to forecast a block. As the lead time \(L\) grows, however, this ordering can invert. A larger \(K\) may ultimately offer more predictable long-term behavior after the system evolves closer to its steady state: with more capacity, the Markov chain dynamics can settle into patterns where blocking occurs less erratically.

\begin{figure}[!ht] 
  \begin{center}
    \includegraphics[width=7.5cm]{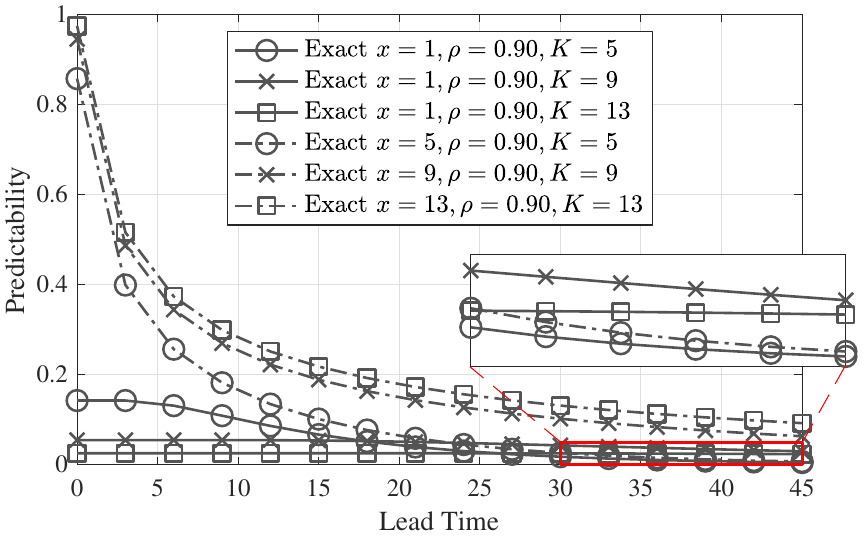}
  \end{center}
  \caption{\bluewraptwo{Predictability of packet loss in a single Geo/Geo/1/K queue under different maximum queue sizes and observed queue lengths, derived using exact numerical calculations in Proposition \ref{proposition:geogeo1loss}.}}
  \label{fig:mm1loss}
\end{figure}

}

\begin{figure}[!ht] 
  \begin{center}
    \includegraphics[width=7.5cm]{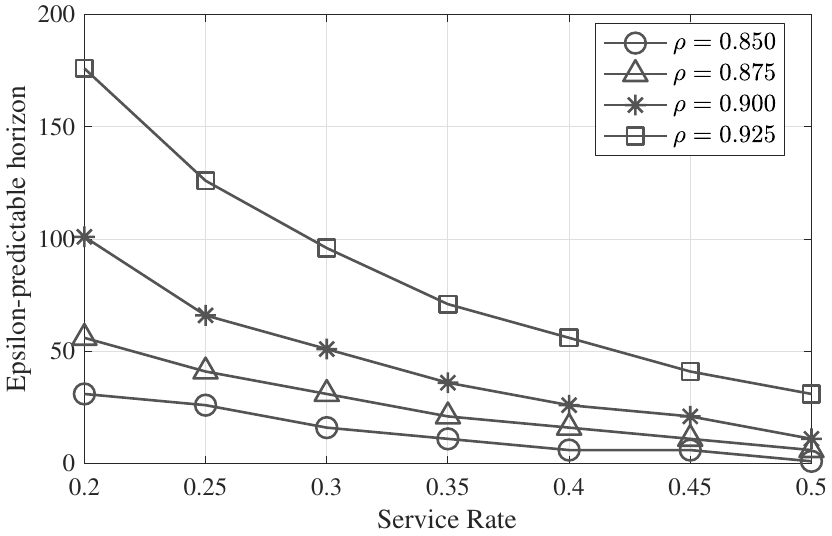}
  \end{center}
  \caption{Epsilon-predictable horizon of Geo/Geo/1/K queuing system, under different configurations and $\epsilon = 0.1$}
  \label{fig:mm1predtime}
\end{figure}

In definition \ref{def:predictabletime}, we introduced the time frame during which the delay predictions remain distinct from the stationary distribution, referred to as epsilon-predictable horizon. 
Figure \ref{fig:mm1predtime} illustrates the epsilon-predictable horizon of different Geo/Geo/1 queues.
For queues with higher utilization, as predictability degrades more slowly, the forecasting system can sample the queue length less frequently and reduce the monitoring overhead.
In a queue with higher arrival rate, but similar service rate, the stationary state distribution is wider and its expected queue length is larger, while the conditional distribution is unchanged.
Therefore, the total variation difference between the forecast and marginal distributions decreases slower as we increase the lead time, resulting in a slower decline in predictability, as illustrated in the figures.

Figure \ref{fig:mm1predtime} illustrates epsilon-predictable horizon for Geo/Geo/1/K queues with $K=50$, highlighting its relationship with both the utilization factor and the service rate.
In this figure, we observe that epsilon-predictable horizon increases in higher utilizations and lower service rates.
Epsilon-predictable horizon is related to the state that results minimum distance between the forecast and marginal distributions, which is close to the expected queue length.

\subsection{Predictability of Multi-Hop Geo/Geo/1 Queues} \label{sec:mh-problem}

In this subsection, we first analyze a multi-hop scenario focusing solely on the impact of the number of hops on system predictability, followed by two multi-hop scenarios with varying levels of observability.

\begin{figure}[!ht] 
  \begin{center}
    \includegraphics[width=7.5cm]{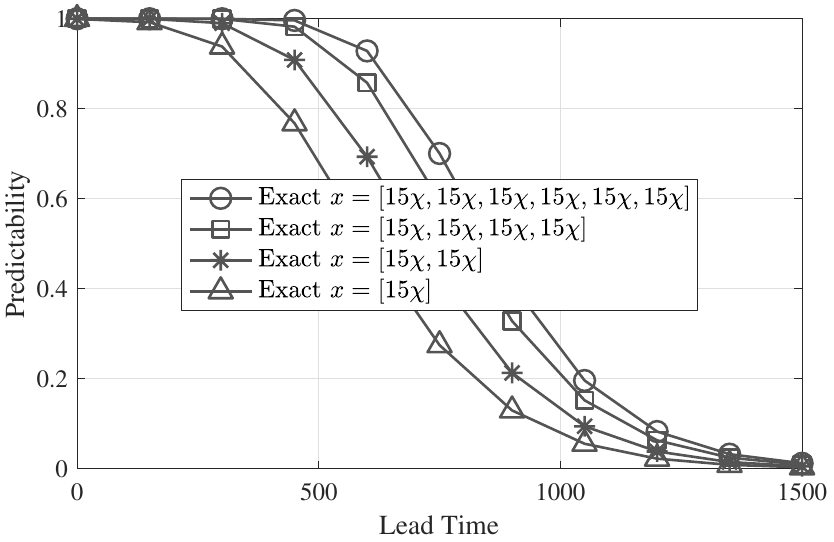}
  \end{center}
  \caption{\bluewrap{Predictability of multi-hop tandem Geo/Geo/1/K queuing systems under different number of hops. All queues have the same service rate of $\mu$ = 0.4 and $\alpha$ = 0.32}}
  \label{fig:mm1multi0}
\end{figure}

\bluewrap{In the first case, shown in Figure \ref{fig:mm1multi0}, we examine tandem Geo/Geo/1/K queues with an increasing number of hops, while keeping the arrival rate ($\alpha = 0.32$) and service rate ($\mu = 0.4$) the same for all queues.  
Counterintuitively, the predictability increases as the number of hops increases. 
This seemingly paradoxical result stems from the fact that we are observing all queues in the system ($x = [15\chi, 15\chi, ...]$) and the target metric is the total sojourn time through the entire system. 
With a single hop, the sojourn time is highly variable. 
However, as the number of hops increases, and each hop's state is known, the total sojourn time becomes the sum of multiple, independent (due to the memoryless property of the Geo/Geo/1 queue) sojourn times.  
By the law of large numbers, the sum of these independent random variables tends towards a more predictable, less variable value. 
Essentially, the individual fluctuations in each queue's sojourn time tend to average out over the entire multi-hop path, leading to a more stable, and therefore more predictable, total sojourn time.
While this reasoning offers a plausible initial explanation, future work should formally investigate and validate this proposed mechanism.}

\begin{figure}[!ht] 
  \begin{center}
    \includegraphics[width=7.5cm]{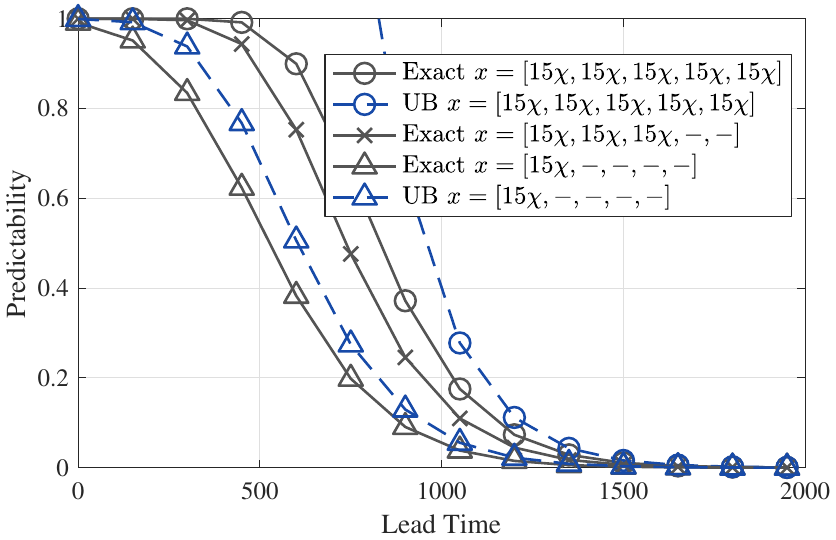}
  \end{center}
  \caption{Predictability of a 5-hop tandem Geo/Geo/1 queuing system under different observability levels. All queues have the same service rate of $\mu=0.4$ and $\alpha=0.32$}
  \label{fig:mm1multi1}
\end{figure}

As illustrated in Figure \ref{fig:mm1multi1}, we evaluate predictability for a system that consists of 5 tandem Geo/Geo/1/K queues, all with the same service rate $\mu=0.4$ and highly congested with the observed state of $15 \chi$.
We consider 3 different configurations:
observations that include all 5 queues, observations including only the first 3 queues, and observations including only the first queue.
We used Equations \ref{eq:multihop1} and \ref{eq:multihop2} to derive exact predictability curves shown via solid lines.
As was expected, predictability decreases as our ability to observe the system declines.
Then the dashed line curves show the upper bounds derived using Lemma \ref{lemma:multihop} and Theorem \ref{theorem:geogeo1pred} (UB).
In the first case where all queues are observable, we derived the bound by summing up the predictability of all hops.
However, when it comes to the case with only one observable queue, the bound equals the predictability of that queue.
This is because if a component is not observable, its predictability is zero and will not contribute to the sum in Equation $\ref{eq:multihopproposition}$.
Therefore, we have $D_n(L) \leq D^{(1)}_n(L)$, which constitutes the upper bound for the partially observable system.

\begin{figure}[!ht] 
  \begin{center}
    \includegraphics[width=7.5cm]{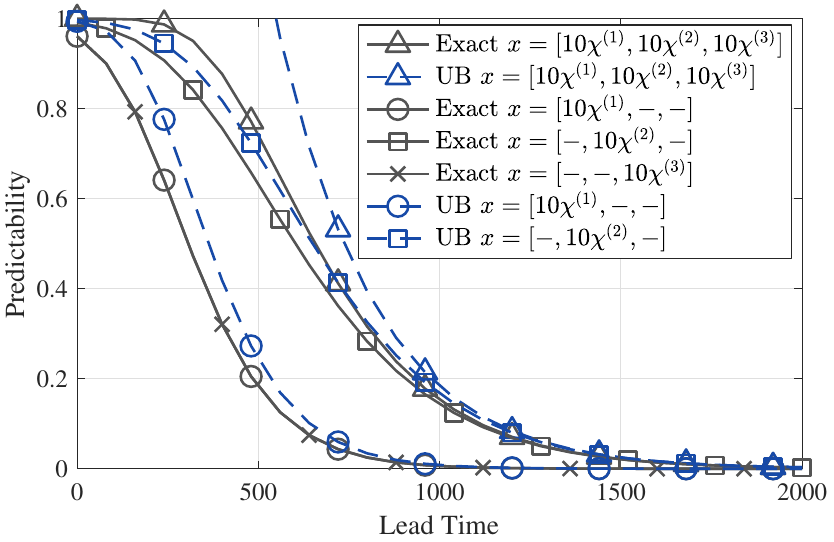}
  \end{center}
  \caption{Predictability of a 3-hop tandem Geo/Geo/1/K queuing system with a bottleneck queue in the middle \bluewrap{and 3 imperfect observation scenarios}. Service rates are $0.4$, $0.38$, and $0.4$ respectively. Arrival rate: $0.34$.}
  \label{fig:mm1multi2}
\end{figure}

The next scenario we analyze in multi-hop networks involves a queue that becomes a bottleneck and how its observability affects predictability. 
Since this queue will contribute more to the end-to-end delay, if it is not observable, we expect the predictability to decrease significantly.
In a setup, we examine 3 tandem Geo/Geo/1/K queues with an arrival rate of $\alpha=0.34$, where the middle queue has a lower service rate, i.e., the service rates are $\mu^{(1)} = 0.4$, $\mu^{(2)} = 0.38$, and $ \mu^{(3)} = 0.4$ respectively.
Suppose $\chi^{(m)}$ as the expected queue length of hop $m$.
All three queues are again highly congested with the observed states
of 10 times their expected queue size.
As shown in Figure \ref{fig:mm1multi2}, we consider 4 different configurations: the observations include all 3 queues, the observations include only the first queue, the observations include only the last queue, and the observation include only the bottleneck queue.
Note that in tandem M/M/1 queues, when the system is in equilibrium, regardless of the service rates, all queues will have the same arrival rate as the first queue \cite{Thomopoulos2012}.
Using that, Equation \ref{eq:multihop1}, and Equation \ref{eq:multihop2}, we derive exact predictability curves shown via solid lines.
The figure indicates that the observation of one of the less utilized queues provides similar benefits in terms of predictability.
However, observing these queues alone results in significantly lower predictability compared to the scenario in which only the bottleneck queue is observable. 
The predictability of the latter case is very close to that of complete observability.
Similar to Figure \ref{fig:mm1multi1}, the dashed lines represent the upper bound derived using Lemma \ref{lemma:multihop} and Theorem \ref{theorem:geogeo1pred} (UB) and it is less tight when the predictability of all hops is summed up.
Note that $\chi^{(1)} = \chi^{(3)}$ due to similar arrival and service rates.

%% file: 6_7_conclusions.tex
\section{Conclusions}

In this paper, we presented a definition of performance predictability for systems with Markovian conditions, with partial and imperfect observability. 
We applied this framework to Geo/Geo/1 queuing models in both single-hop and multi-hop contexts, \bluewrap{as well as to a random-walk-based model representing the connection state of the vehicles traversing in a \gls{V2X} environment.} We illustrated the practical implications of studying predictability in these systems. Our evaluations highlight the effectiveness of the proposed definitions in modeling crucial trade-offs in future communication systems, \bluewrap{including V2X applications,} incorporating \gls{QoS} predictors. For example, we illustrated the overhead-predictability tradeoff and demonstrated how a lack of observability can deteriorate predictability.

In multi-hop networking, we showed both exact and upper bounds on the predictability, with reduced monitoring resources compared to observing all queues, showed monitoring the bottleneck queue is crucial for maintaining predictability.
Additionally, state aggregation is a viable option to manage the overhead of observations in large-scale networks. 

The behavior of the underlying Markov processes greatly impacts the predictability of the system. 
Drawing on the literature on Markov chain mixing times, including spectral gap analysis, provides insights into how random system conditions transition.
We examined the connection between system predictability and Markov chain behavior by employing an upper limit on predictability derived from spectral analysis methods. 
Future research can explore conductance analysis to enhance our understanding of Markov chain dynamics.
A possible direction for future research is to identify or detect hidden Markov models in the environment with which the communication network interacts. 
Incorporating these models into predictability analysis will offer insights into the feasibility and timescales of \gls{QoS} predictions.

In future communication systems where performance prediction is essential, it is important to improve predictability efficiently.
This involves enhancing observability and directing monitoring efforts towards critical points within the network, such as performance bottleneck nodes or queues. 
Additionally, it is necessary to identify whether the environment or application's behavior is predictable and subsequently learn it for performance prediction. 
Incorporating these insights can lead to more efficient communication systems.


%% file: 8_appendix.tex
\begin{appendices}
\numberwithin{equation}{section}

\section{Proof of Theorem \ref{theorem:upperbound}} \label{app:upperbound}
In this proof, our approach is to bound \gls{TV} distance using chi-squared distance.
For arbitrary probability distributions $p$ and $q$, according to Pinsker's inequality, total variation distance is upper bounded as
\begin{equation} \label{eq:chisquaredineq}
    \norm{p - q}_{\text{TV}} \leq \sqrt{ \frac{1}{2} \chi^2(p, q)}
\end{equation}
where $\chi^2(p, q)$ denotes chi-squared distance and is obtained
\begin{equation}
    \chi^2(p, q) = \sum_{z \in \mathcal{Z}} \frac{(p(z) - q(z))^2}{q(z)}.
\end{equation}
Therefore, we begin by bounding chi-squared distance between forecast and marginal distributions which can be derived as
\begin{multline} \label{eq:chisquared}
\chi^2(\Pr(Z_{n+L}\mid X_{n} = x),  \Pr(Z_{n+L})) = \\
\sum_{z \in \mathcal{Z}} \frac{\left(\sum_{y \in \mathcal{X}} \left(P^L(x, y) - \pi(y)\right) r_y(z)\right)^2}{Pr(Z_{n+L})}.
\end{multline}
We use the following equations in our proof from the spectral theorem in Markov chain mixing time \cite{levin2017markov}:
\begin{align}
    &\frac{P^{L}(x,y)}{\pi(y)} = 1 + \sum_{j=2}^{K} f_j(x) f_j(y)\lambda_{j}^{L}, \\
    &\sum_{i=2}^n f_i(x) f_i\left(x^{\prime}\right) =\frac{\mathbf{1}\left(x=x^{\prime}\right)}{\pi(x)}-1.
\end{align}
First, we substitute $P^L(x, y) - \pi(y)$ from above in \ref{eq:chisquared} and obtain
\begin{multline}
    \chi^2(\Pr(Z_{n+L}\mid X_{n} = x),  \Pr(Z_{n+L})) = \\
    \sum_z \frac{\left( \sum_y\sum_{j=2}^{|\mathcal{X}|} f_j(x) f_j(y) \lambda_{j}^{L}\pi(y)r_y(z) \right)^{2}}{\sum_y{\pi(y)r_y(z)}}.
\end{multline}
Then we apply Cauchy-Schwarz inequality as follows
\begin{multline}
    \left( \sum_y\sum_{j=2}^{|\mathcal{X}|} f_j(x) f_j(y) \lambda_{j}^{L}\pi(y)r_y(z) \right)^{2} = \\
    \left( \sum_{j=2}^{|\mathcal{X}|} f_j(x) \lambda_{j}^{L} \sum_y f_j(y) \pi(y)r_y(z) \right)^{2} \leq \left( \sum_{j=2}^{|\mathcal{X}|} f^2_j(x) \lambda^{2L}_{j}\right) \cdot \\
     \left( \sum_{j=2}^{|\mathcal{X}|} \sum_{y,y'} f_j(y) \pi(y)r_y(z) f_j(y') \pi(y')r_{y'}(z) \right) \\
    \leq \lambda^{2L}_{*} \left( \sum_{j=2}^{|\mathcal{X}|} f^2_j(x) \right) \cdot \\
    \left( \sum_{y,y'} \pi(y)r_y(z) \pi(y')r_{y'}(z)\sum_{j=2}^{|\mathcal{X}|} f_j(y)f_j(y') \right) \\
    = \lambda^{2L}_{*} \left( \frac{1}{\pi(x)} - 1 \right) \cdot \\ 
    \left( \sum_{y,y'} \pi(y)r_y(z) \pi(y')r_{y'}(z) \left( \frac{\mathbf{1}(y=y')}{\pi(y)} - 1 \right) \right),
\end{multline}
where $\mathbf{1}()$ denotes the unit function.
\begin{multline}
    \chi^2(\Pr(Z_{n+L}\mid X_{n} = x),  \Pr(Z_{n+L})) \leq \lambda^{2L}_{*} \left( \frac{1}{\pi(x)} - 1 \right) \cdot \\
    \sum_z \frac{\sum_{y,y'} \pi(y)r_y(z) \pi(y')r_{y'}(z) \left( \frac{\mathbf{1}(y=y')}{\pi(y)} - 1 \right)}{\sum_y{\pi(y)r_y(z)}}
\end{multline}
and for the right term we simplify
\begin{multline}
    \sum_z \frac{\sum_{y,y'} \pi(y)r_y(z) \pi(y')r_{y'}(z) \left( \frac{\mathbf{1}(y=y')}{\pi(y)} - 1 \right)}{\sum_y{\pi(y)r_y(z)}} = \\
    \sum_z \frac{\sum_{y} \pi(y)r^{2}_y(z)}{\sum_y{\pi(y)r_y(z)}} - 1
\end{multline}

\begin{multline}
    \chi^2(\Pr(Z_{n+L}\mid X_{n} = x),  \Pr(Z_{n+L})) \leq \\
    \lambda^{2L}_{*} \left(\frac{1}{\pi(x)} - 1\right)\left( \sum_z \frac{\sum_{y} \pi(y)r^{2}_y(z)}{\sum_y{\pi(y)r_y(z)}} - 1\right),
\end{multline}
Incorporate this into \ref{eq:chisquaredineq} we get
\begin{multline}
    D_n(L) = \norm{\Pr(Z_{n+L}\mid X_{n} = x) - \Pr(Z_{n+L}))}_{\text{TV}} \leq \\
    \frac{1}{\sqrt{2}}\lambda^{L}_{*} \left(\frac{1}{\pi(x)} - 1\right)^{1/2}\left( \sum_z \frac{\sum_{y} \pi(y)r^{2}_y(z)}{\sum_y{\pi(y)r_y(z)}} - 1\right)^{1/2}.
\end{multline}

\section{Proof of Proposition \ref{proposition:geogeo1approx}} \label{app:geogeo1approx}

We start from Theorem \ref{theorem:geogeo1pred}, Equation \ref{eq:mm1pred} where in the inner sum over $k$, the term $\gamma^L_{k}$ converges to zero, while other terms remain nonzero, but only when $L$ is large, and except when $k/(K+1)$ is small.
Therefore, we simplify the rest of the terms for the case when $k/(K+1)$ is small.
As for the $\sin$ terms, as $k/(K+1)$ is small, we approximate 
\begin{multline}
   \sin \left(\frac{(x+1) k \pi}{K+1}\right) \approx \sin \left(\frac{x k \pi}{K+1}\right),\\
   \sin \left(\frac{(y+1) k \pi}{K+1}\right) \approx \sin \left(\frac{y k \pi}{K+1}\right).
\end{multline}
We can simplify Theorem \ref{theorem:geogeo1pred} as
\begin{multline}
    D_n(L) = \frac{1}{2}\sum_{z} \biggl| \sum_{y=1}^{K} \frac{2}{K+1}\beta^{\frac{y-x}{2}} (1-\sqrt{\beta})^2 \\ \sum_{k=1}^{K} \frac{\left(\alpha \mu+\bar{\alpha}\bar{\mu}+ 2 \sqrt{\alpha \mu\bar{\alpha}\bar{\mu}} \cos (\frac{k \pi}{K+1})\right)^L}{\left[1-2 \sqrt{\beta} \cos \left(\frac{k \pi}{K+1}\right)+\beta\right]} \\
    \sin \left(\frac{x k \pi}{K+1}\right) \sin \left(\frac{y k \pi}{K+1}\right) \text{NB}(z; y,\mu) \biggr|.
\end{multline}
We continue the approximation with the summation over $y$:
\begin{multline}
    \sum_{y=1}^{K} \beta^{y/2}\sin{ \left( \frac{y k \pi}{K+1} \right) } \text{NB}(z; y,\mu) \approx \\
   \sqrt{\beta} \frac{\sin{\left(\frac{k\pi}{K+1}\right)}}{1-2\sqrt{\beta}\cos{\left(\frac{k \pi}{K+1}\right)} + \beta} \text{NB}(z; 1,\mu).
\end{multline}
Since $\beta^{y/2}$ decays exponentially, the dominant contributions come from the lower values of $y$.
Therefore, we substituted the NB term with $\text{NB}(z; 1,\mu)$.
The remaining terms inside the summation are easy to show that for large $K$, and $\beta < 1$, can be approximated accordingly.

Next we simplyfy $\gamma_k^L$ term using the binomial approximation for large $L$:
\begin{multline*}
   (a + b \cos(c))^L \approx (a + b)^L \left(1 - \frac{b c^2}{2(a + b)}\right)^L \approx \\
   (a + b)^L e^{-L \frac{b c^2}{2(a + b)}}.
\end{multline*}
Substitute $\gamma_k^L$ with above approximation will give us:
\begin{multline}
   \left(\alpha \mu+\bar{\alpha}\bar{\mu}+ 2 \sqrt{\alpha \mu\bar{\alpha}\bar{\mu}} \cos (\frac{k \pi}{K+1})\right)^L \approx \\ 
   \left(\alpha \mu+\bar{\alpha}\bar{\mu}+ 2 \sqrt{\alpha \mu\bar{\alpha}\bar{\mu}}\right)^L e^{- L \kappa (k\pi/(K+1))^2},
\end{multline}
where
\begin{equation*}
   \kappa = \frac{\sqrt{\alpha\mu\bar{\alpha}\bar{\mu}}}{\alpha\mu + \bar{\alpha} \bar{\mu}+2\sqrt{\alpha\mu\bar{\alpha}\bar{\mu}}}.
\end{equation*}
Next, since $K$ is large, we convert the sum $\sum_k$ to integral by substituting $k\pi/(K+1)$ with $r$ as
\begin{multline}
\sum_k  \frac{\sin{\left(\frac{k\pi}{K+1}\right)} \sin{\left(\frac{k x \pi}{K+1}\right)} }{1-2\sqrt{\beta}\cos{\left(\frac{k \pi}{K+1}\right)} + \beta} e^{- L \kappa (k\pi/(K+1))^2} =\\
\int_{0}^{\pi} \frac{\sin{r}\sin{x r}}{(1-2\sqrt{\beta}\cos{r}+\beta)^2} e^{-L \kappa r^2} dr.
\end{multline}
Since the integral and the rest of the terms are positive, we remove the absolute value operator and move $\sum_{z}$ inside to the only term dependent on $z$, which is the NB term and since it is a pmf, the summation results $1$.

Finally, by combining all approximations we reach the final term as
\begin{multline}
    D_n(L) \approx \frac{\beta^{\frac{1-x_n}{2}}}{\pi}\left(1-\sqrt{\beta}\right)^2\left(\alpha\mu + \bar{\alpha} \bar{\mu} + 2\sqrt{\alpha\mu\bar{\alpha}\bar{\mu}}\right)^{L} \\ \int_{0}^{\pi} \frac{\sin{r}\sin{x r}}{(1-2\sqrt{\beta}\cos{r}+\beta)^2} e^{-L \kappa r^2} dr.
\end{multline}

\section{\bluewrap{Random Walk Parameter Extraction}}  \label{app:randomwalk}

\bluewrap{We analyzed a real-world dataset comprising LTE measurements collected from multiple CSV files introduced in \cite{Raca2018}, each containing time-series entries for attributes such as \texttt{NetworkMode}, \texttt{CQI}, \texttt{Speed}, and \texttt{DL\_bitrate}. For each file, we retained rows that corresponded to operator \texttt{A}, used the \texttt{LTE} mode, and featured valid \gls{CQI} readings. Downlink bitrates (originally in kbps) were rounded to the nearest discrete value in Mbps. After discarding missing or insufficient data, we combined the cleaned results across all files for further analysis.}

\begin{figure}
	\centering
	\begin{subfigure}{1\linewidth}
            \centering
		\includegraphics[width=7.5cm]{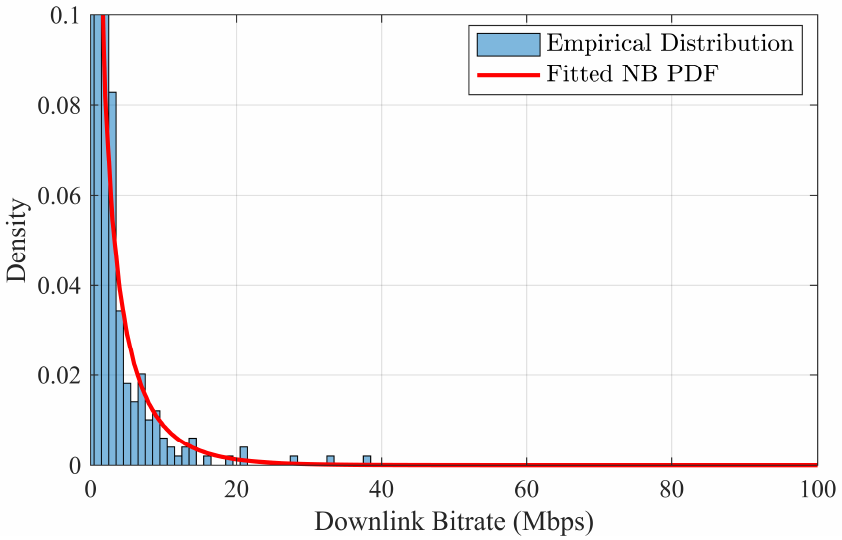}
		\caption{\bluewrap{\gls{CQI} = $3$}}
            \label{fig:nbfita}
	\end{subfigure}
        \begin{subfigure}{1\linewidth}
            \centering
		\includegraphics[width=7.5cm]{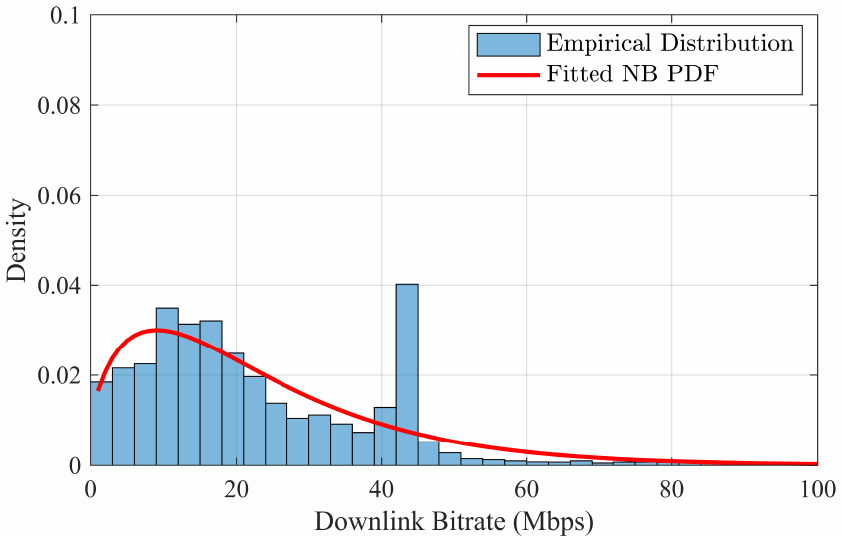}
		\caption{\bluewrap{\gls{CQI} = $12$}}
            \label{fig:nbfitb}
	\end{subfigure}
	\caption{\bluewrap{Empirical vs. fitted negative binomial distributions for two \gls{CQI} values and measured downlink LTE throughput in Mbps from dataset introduced in \cite{Raca2018}.}}
	\label{fig:nbfit}
\end{figure}
\begin{figure}
	\centering
	\begin{subfigure}{1\linewidth}
            \centering
		\includegraphics[width=7.5cm]{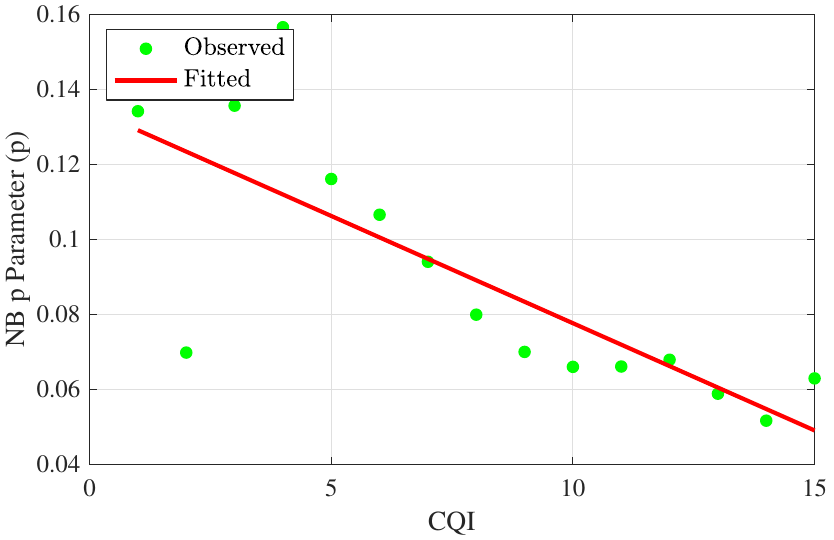}
		\caption{\bluewrap{Negative binomial parameter $p$ estimation via $\alpha_p \,\mathrm{CQI} + \beta_p$}}
            \label{fig:nbrega}
	\end{subfigure}
        \begin{subfigure}{1\linewidth}
            \centering
		\includegraphics[width=7.5cm]{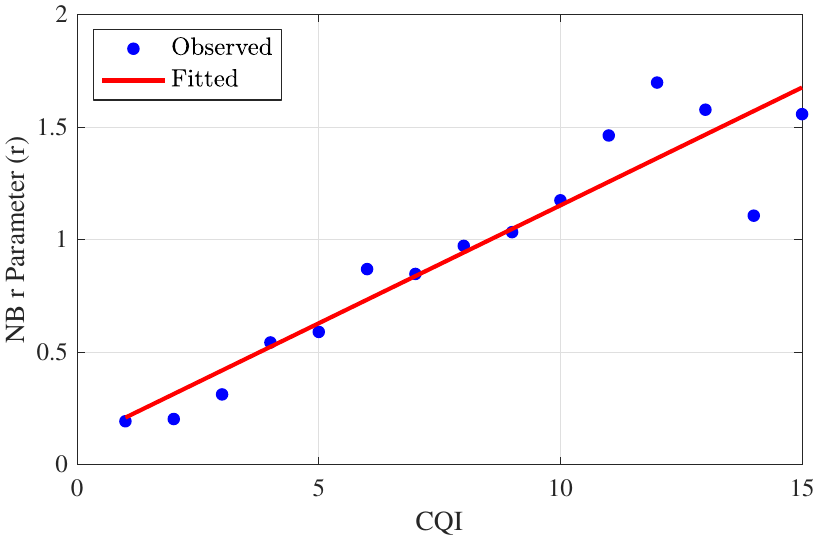}
		\caption{\bluewrap{Negative binomial parameter $r$ estimation via $\alpha_r \,\mathrm{CQI} + \beta_r$}}
            \label{fig:nbregb}
	\end{subfigure}
	\caption{\bluewrap{Linear regression results to determine the relationship between the parameters of the negative binomial distribution and different \gls{CQI} values.}}
	\label{fig:nbreg}
\end{figure}

\bluewrap{Next, we fit a Negative Binomial (NB) distribution to the downlink throughput for each \gls{CQI} level \(\{1,\dots,15\}\) as we illustrate two examples in Figure \ref{fig:nbfit}. Whenever a \gls{CQI} level had sufficiently many samples (e.g., more than 10 valid observations), we used the method of moments to estimate the NB parameters \(\bigl(r, p\bigr)\). We then performed a linear regression across the \gls{CQI} levels to relate the estimated parameters \(\{\,r,\,p\}\) to \gls{CQI} in the form 
\[
r(\mathrm{CQI}) \;=\; \alpha_r \,\mathrm{CQI} + \beta_r,
\quad
p(\mathrm{CQI}) \;=\; \alpha_p \,\mathrm{CQI} + \beta_p,
\]
and saved both the raw estimates and the regression results for subsequent simulation.
The regressions results are depicted in Figure \ref{fig:nbreg}.}

\bluewrap{To capture how \gls{CQI} evolves over time, we employed a lazy random walk model on a finite state space \(\{1,\dots,15\}\), in which the CQI remains unchanged between consecutive time steps with probability \(p\), and transitions to adjacent states with probability \(\frac{1-p}{2}\). We estimated \(p\) empirically by counting how often \(X_{n+1} = X_n\) in the dataset. To distinguish between vehicular (mobile) and near-static connections, we applied a speed threshold (e.g., 20\,km/h) and computed separate values of \(p\). In the dataset at hand, vehicular connections yielded \(p \approx 0.6\), whereas near-static connections yielded \(p \approx 0.9\).
These results reflect the differing degrees of CQI fluctuation in mobile versus almost-static environments.
Thus, with the estimated parameters, we construct the conditional distribution and define the corresponding Markov chain, enabling the application of the predictability framework.}

\end{appendices}